\documentclass{optica-article}

\journal{opticajournal} % for journals or Optica Open

\articletype{Research Article}

\usepackage{lineno}
\usepackage{mathtools}
\begin{document}
\newcommand{\R}{{\rm I\!R}}
\title{Scattering graph method for 3D radiative transfer}

\author{Antti Mikkonen\authormark{1,*}, Anssi Koskinen\authormark{1,2}, Johanna Tamminen\authormark{1} and Hannakaisa Lindqvist\authormark{3} }

\address{\authormark{1}Finnish Meteorological Institute, Dynamicum,
Erik Palménin aukio 1, FI-00560, Helsinki, Finland\\
\authormark{2}University of Helsinki, Pietari Kalmin katu 5, FI-00560 Helsinki, Finland\\
\authormark{3}Finnish Meteorological Institute, Arctic Space Centre, Tähteläntie 62, FI-99600 Sodankylä, Finland}

\email{\authormark{*}antti.mikkonen@fmi.fi} %% email address is required; see note below about the corresponding author designation

% use {asbstract*} to suppress the copyright line. Copyright information will be added in production

\begin{abstract*} 
A novel method for monochromatic scalar 3D radiative transfer, designed primarily for modeling remote sensing imaging, is presented. For simulating an observation of an imaging satellite instrument, the method uses a heuristic scattering coupling function to model the inter-pixel scattering of radiation, which is represented with a graph. The GPU-capable code implementation of the method, TURSCA, was validated against two established 3D RT models, Siro and SHDOM with relative agreement at 3\% and 6\%, respectively. The capabilities of TURSCA in modeling a satellite observation of an emission plume are examined. The presented method opens up new avenues of research, especially in satellite-based remote sensing of atmospheres.%Finding the optimal scattering coupling parameter was shown to be a non-trivial task and it necessitates further research.%The abstract should be limited to approximately 100 words. If the work of another author is cited in the abstract, that citation is written out as, for example, Opt. Express {\bfseries 32}, 32643 (2024), and a separate citation should be included in the body of the text. The first reference cited in the main text must be [1].There is no need to include numbers, bullets, or lists inside the abstract. Do not add the licensing or copyright statement at submission.

% Using this scattering graph, the optical paths within the computation medium can be modelled and their combined contribution to the observed satellite radiances can be calculated simply by solving a linear system defined by the flux matrix. The flux matrix is composed of discrete-ordinates-like integrals of the scattering phase functions in an irregular grid of the graph nodes and the Beer-Lambert-Bouguer attenuation on the graph edges.

\end{abstract*}

%%%%%%%%%%%%%%%%%%%%%%%%%%  body  %%%%%%%%%%%%%%%%%%%%%%%%%%

\section{Introduction}

Atmospheric remote sensing is the process of inferring information about the planetary atmosphere based on observations of electromagnetic radiation which it has interacted with, and this information is extracted from the observations with a retrieval algorithm. At the core of atmospheric remote sensing retrieval algorithms is a radiative transfer (RT) model, which simulates the phenomena associated with the propagation of radiation in a medium, such as absorption, scattering and emission, in the atmosphere. 

In the context of inferring trace gas concentrations from space-based spectral measurements, for example, the Atmospheric CO$_2$ Observations from Space (ACOS) algorithm \cite{amt-5-99-2012-odell}, used in retrieving total column of carbon dioxide (CO$_2$) from NASA's Orbiting Carbon Observatory -2 (OCO-2) satellite observations, utilizes a modified version of LInearized Discrete Ordinate Radiative Transfer (LIDORT) \cite{spurr2008lidort} and the methane column retrieval algorithm \cite{amt-9-5423-2016-hu} employs LINTRAN \cite{SCHEPERS2014347} to process TROPOspheric Monitoring Instrument (TROPOMI) observations. In both of these RT models, the atmosphere is assumed to be one-dimensional, with the only variation being altitudal. This is a good assumption when the ground pixel sizes are in the scale of several km$^2$ as is the case with OCO-2 and TROPOMI. In addition, OCO-2 and TROPOMI are both in a solar synchronous orbit with their equatorial crossing times close to local noon, which cause the observed solar zenith angles (SZA) be relatively small, further supporting the assumption of 1D atmosphere and yielding less atmospheric scattering compared to larger SZAs, which can complicate the retrieval. In the vicinity of clouds, however, the RT modeling in a 3D atmosphere is required for OCO-2 and TROPOMI \cite{amt-18-73-2025-trees,amt-16-2145-2023}. A nearby cloud may cast a shadow onto the pixel to be retrieved, causing a deviation from the assumed 1D atmosphere, resulting into biases in the retrieved trace gas column concentrations. These cloud 3D effects are the most prominent, but airborne aerosols and gas concentration differences could cause 3D effects as well.

These 3D effects likely become more prominent when retrieving atmospheric information from high-resolution imaging spectrometer satellite instruments, such as GHGSat \cite{amt-14-2127-2021-jervis}, EMIT \cite{nasa-emit} and PRISMA \cite{8518512-prisma}. In nadir-viewing passive remote sensing, when the ground pixel sizes are in the scale of thousands of m$^2$, the incident solar radiation will be affected by atmospheric columns in several adjacent pixels due to their small size. Atmospheric scattering further increases the radiative interaction between the columns. This effect between the pixels is also increased with the increasing SZA. In high latitudes, where SZAs are consistently large, these adjacent pixel interactions arise even in the km$^2$ size scale. Additionally, large viewing zenith angles (VZA) strengthen the radiative interaction with neighbouring pixels. A case with large SZAs and VZAs could be Sentinel-4 imaging spectrometer observations of the high latitudes from the geostationary orbit \cite{10.1117/12.2304099-sentinel4}, OCO-2 observations of the sunglint reflected from the Arctic ocean, or continuous observation of a single point on the Earth's surface throughout the whole satellite overpass (i.e. target mode observation). Therefore, to be as realistic as possible, the RT problem in these cases should not be considered as individual pixels with 1D RT, but instead a full 3D RT should be calculated.

There are two main approaches for 3D RT: discrete ordinates and Monte Carlo. These two different approaches have been verified to give consistent results \cite{pincus-2009-mc-shdom,jones-2018-mc-shdom}. Theoretically these methods are quite general, but their computational implementations can be based on widely different assumptions. These 3D RT codes are thoroughly intercompared by the International Intercomparison of 3D Radiation Codes (I3RC) project \cite{i3rc}. %For more information regarding atmospheric 3D RT, refer to \cite{marshak20053d, mayer-rt-2009}.

With discrete ordinates the radiation field is solved at each volume element into discrete zenith and azimuthal angular grid. This systematic approach is conceptually simple, but sets constraints on medium geometries and it can be prohibitively expensive to compute if a small subset of the radiation field is of interest, as is the case in satellite remote sensing. The discrete ordinates approaches can be used conjointly with other numerical methods to tackle the 3D RT problem. The Spherical Harmonic Discrete Ordinates Method (SHDOM) combines modeling of source and scattering functions using spherical harmonics with spatial integration using discrete ordinates \cite{shdom}. The Discrete Anisotropic Radiative Transfer (DART) model employs ray-tracing and discrete ordinates \cite{GRAU2013149-DART}. 

In Monte Carlo RT (e.g. \cite{govaerts1998raytran}), photon packets are traced through the medium and are scattered into different directions sampled from the scattering phase functions. It is straightforward to incorporate different phenomena into the computation, such as polarization or refraction. However, Monte Carlo requires simulating millions of photon packets, which can be slow to compute and is non-deterministic due to stochastic nature of scattering. Code implementations for 3D Monte Carlo RT are, for example, I3RC community Monte Carlo \cite{pincus-2009-mc-shdom}, LargE-Scale Simulation framework (LESS) \cite{QI2019695-LESS} and MONKI (Monte Carlo KNMI) \cite{amt-18-73-2025-trees}.

Due to the heavy computational load present in the RT calculations, graphical processing units (GPU) have been of increasing use. GPUs have been primarily designed for rapidly visualizing computer graphics, but with the advent of CUDA\cite{luebke-cuda-2008} and OpenCL\cite{munshi-opencl-2009}, they can be used for more general computation also. The main difference between GPUs and CPUs are that GPUs control thousands of computation threads compared to CPU's tens of threads, but the GPU threads operate on single-instruction-multiple-data (SIMD) paradigm, where maximum computation efficiency is attained when homogenous calculations are carried out on heterogenous data. This makes the GPUs especially suitable for Monte Carlo RT models\cite{RAMON201989,HUANG20112207,bian-2022-3drtmc,alerstam-2008}, of which there are several implementations, but also the discrete ordinates method benefits from the GPU implementation\cite{EFREMENKO20143079}.

In this work a novel methodology for 3D RT is presented. The scattering graph method is a way to model monochromatic scalar radiation fields observed by a remote sensing instrument in an absorbing and scattering medium, which is as flexibly defined as in Monte Carlo models, while being faster and deterministic. The method aims to enable new approaches on satellite-based atmospheric remote sensing retrieval algorithm development. The software implementation of the method, Transmittance calcUlator with Radiative Scattering Coupling Approach (TURSCA), can be used on both CPUs and GPUs for simulating remote sensing observations from satellite-based imagers.

The rest of the paper is structured as follows: first the scattering graph approach is examined from a theoretical point of view, and then the structure of the GPU-capable computational solver TURSCA is explained. After that, TURSCA is validated against two other RT models in scenes with large SZAs and VZAs. Finally, the effect of scattering coupling on the simulated transmittances is analyzed and overall conclusions of the paper are outlined.
\section{Theoretical basis of scattering graph RT solver}\label{sec:theory}
The radiative transfer equation (RTE) \cite{ishimaru1978wave}, in domain $\Omega \subset \R^3$ is 
\begin{equation}
\begin{split}
\hat{s}\cdot \nabla \phi(r,\hat{s}) + (\mu_\mathrm{s}(r) + \mu_\mathrm{a}(r))\phi(r,\hat{s}) = \epsilon(r,\hat{s}) + \mu_\mathrm{s}(r)\int_{S_{2}} \Theta(r,\hat{s} \cdot \hat{s}')\phi(r,\hat{s}')\mathrm{d}\hat{s}',\:\:& r\in\Omega \\
\phi(r,\hat{s})  = \phi_0(r,\hat{s}),\:\:&  r \in\partial\Omega
\end{split}
\label{RTE}
\end{equation}
where $\phi(r,\hat{s})$ is the radiance at position $r\in\Omega$ into direction $\hat{s}\in\R^3, \|\hat{s}\|_2=1$, $\mu_s(r)$ is the scattering coefficient at $r$, $\mu_a(r)$ is the absorption coefficient at $r$, $\epsilon(r,\hat{s})$ is the emissivity at the position $r$ into the direction $\hat{s}$, $S_2$ is the surface of a 2-sphere and $\Theta(r,\hat{s}\cdot\hat{s}')$ is the scattering phase function at position $r$ from the direction $\hat{s}'$ into the direction $\hat{s}$.

In many remote sensing and imaging applications, not the full solution (i.e. for every $r\in\Omega$) of $\phi(r,\hat{s})$ is of interest. For some observation instrument, we want to solve $\phi(r,\hat{s})$ such that $\hat{s} \in A_{S_2} \subset S_2$ and $r \in A_\Omega \subset \partial\Omega$ if the instrument is outside the domain or at a single point $r = r_{\mathrm{instrument}}\in\Omega$ if the instrument resides within the domain. In either case, we can select $N_{\mathrm{LOS}}$ pairs $(r_0^i,\hat{s}_0^i)$, $i = 0,1,...,N_\mathrm{LOS}$, such that $-\hat{s}_0^i \in A_{S_2}$ and $r_0^i \in A_\Omega$ or $r_0^i = r_{\mathrm{instrument}}$ for every $i$.

Each of the pairs $(r_0^i,\hat{s}_0^i)$ define a \textit{line-of-sight} (LOS) along which a path can be traced through the domain starting at $r_0^i$ into the direction $\hat{s}_0^i$. A path is a collection of points $r_j^i \in \R^3$, $j = 0,1...,N_\mathrm{step}^i$, such that $(r_{j+1}^i - r_{j}^i) / \|r_{j+1}^i - r_{j}^i\|_2 := \hat{s}_j^i$. In the case of no refraction in the domain $\hat{s}_j^i = \hat{s}_0^i$ for all $j$. The points $r_j^i$ along a particular line-of-sight can be selected at constant distance intervals, at each atmospheric layer or by using the mean free path, $1/\mu_s$, as the step length. The choice of the path creation strategy depends on the medium properties, size scales so that all the phenomena of interest are simulated and the line-of-sight positioning so that path point spacing is ideally isotropic among all the path points within the domain.

Let's consider two different points $r_j^i$ and $r_l^k, k = 0,1,...,N_\mathrm{LOS}, l = 0,1...,N_\mathrm{step}^k$. In general case, the radiance at point $r_j^i$ has some effect on the radiance at $r_l^k$, or in more mathematical terms, 
\begin{equation}
    \frac{\partial\phi(r_l^k,\hat{s})}{\partial\phi(r_j^i,\hat{s}')} \neq 0,
    \label{eq:radiance_effectuality}
\end{equation}
with some $\hat{s},\hat{s}'\in S_2$. If the integral term in Eq. \ref{RTE} is disregarded (i.e. $\Theta(\hat{s},\hat{s}') = 0$ for every $\hat{s}$ and $\hat{s}'$), then the RTE becomes an inhomogeneous first-order linear PDE which can be solved analytically. Moreover, directions $\hat{s}$ can be solved independently so the partial derivative in Eq. \ref{eq:radiance_effectuality} is zero if $\hat{s} \neq \hat{s}'$. In the case of $\hat{s} = \hat{s}'$, Eq. \ref{eq:radiance_effectuality} becomes
\begin{equation}
    \frac{\partial\phi(r_l^k,\hat{s})}{\partial\phi(r_j^i,\hat{s})} = \frac{\mu_e(r_l^k)}{\mu_e(r_j^i)}\exp\left(-\int_{r_j^i}^{r_l^k}\mu_e(z)\mathrm{d}z\right),
\end{equation}
in which $\mu_e = \mu_s + \mu_a$. That is to say the effect of radiance at the point $r_j^i$ affects the radiance at point $r_l^k$ in accordance with the well-known Beer-Lambert-Bouguer attenuation law.

Examining the effect of scattering between the points $r_j^i$ and $r_l^k$ is a non-trivial task. At this point it must be noted that for this question to be physically meaningful, scattering between the points $r_j^i$ and $r_l^k$ should be possible -- with low values of $\mu_s$, scattering is possible, but it can be insignificant, which poses merely computational nuisance, whereas with very high values of $\mu_s$, the mean free path ($1/\mu_s$) can be much shorter than the distance between $r_j^i$ and $r_l^k$, indicating that scattering between these two points is not only minimal, but also unphysical. Therefore, extra care is needed when analyzing highly scattering media from this point-of-view.

One approach is to use a heuristic function to estimate the Eq. \ref{eq:radiance_effectuality}. Let's define a \textit{scattering coupling} function $C: \R^3 \times \R^3 \rightarrow \R$, which gives some numerical value how radiance in one point affects radiance in another point. The scattering coupling function is now defined to be
\begin{equation}\label{eq:scatt_coup}
    C(r,r') = \frac{\sqrt{\mu_s(r) \cdot \mu_s(r')}}{\|r' - r\|_2},
\end{equation}
but alternative definitions with generalized mean $(\frac{1}{2}(\mu_s(r)^b + \mu_s(r')^b))^{1/b}$, $b\in\R$, $l^p$ norm $\|r'-r\|_p$, $p\geq1$, and scattering phase functions could be explored in further studies. Once the $C(r_j^i,r_l^k)$ has been computed for all the pairs, disregarding the $r = r'$ case, we can select a subset of these pairs for which the scattering coupling is above some threshold value $c_\mathrm{min}$. The magnitude of $c_\mathrm{min}$ should be selected with the function $C$ in mind so that the effect of $\phi(r_j^i,\hat{s}')$ onto $\phi(r_l^k,\hat{s})$ will be insignificant. In other words, $c_\mathrm{min}$ needs to be small enough so that all desired scattering phenomena are captured, but large enough so that unnecessary computational burden is avoided. These couplings form a graph $G(R,L)$, where nodes $R = \{r_j^i \;|\; i = 0,1,...,N_\mathrm{LOS}, j = 0,1...,N_\mathrm{step}^i\}$ and edges $L = \{(r_j^i,r_l^k) \;|\; C(r_j^i,r_l^k) > c_\mathrm{min}\}$. The graph $G(R,L)$ is now called a \textit{scattering graph}. 

The physical interpretation of the scattering graph is that it describes all the possible paths a scattered light beam can take in the domain by moving from one node to another along the edges. Comparatively in a Monte Carlo model, these paths are traced by randomly sampling the scattering point from optical depth along a path and interpreting the scattering phase function as a probability distribution function to select a scattering direction. Even though there are infinite number of paths the scattered light can traverse within the scattering graph, the radiative energy is attenuated along the edge between the two nodes (i.e. multiplied with a factor strictly less than one) and therefore the combined contribution of all the paths can be calculated. Using a matrix $\mathbf{F}$ to represent how much radiative energy from one node traversing to a neighbouring node can scatter toward its neighbour (which can be the initial node as well), the total radiative fluxes between the nodes caused by incident radiation $S$ is 
\begin{equation}\label{eq:fs_sum}
    \varphi = S + \mathbf{F}S + \mathbf{F}^2S + \mathbf{F}^3S...
\end{equation}
which is a geometric series of matrices representing the radiative flux after each successive scattering event. This matrix $\mathbf{F}$ is now called the \textit{flux matrix}. In the terms of graph theory, the matrix $\textbf{F}$ is now an adjacency matrix of the line graph of the scattering graph.

Using the scattering graph, the flux matrix is constructed. Let $a,b \in 0,1,...,\#L := N_\mathrm{coupling}$, $L_a$ and $L_b$ are the $a$'th and $b$'th element of $L$ and flux matrix $\mathbf{F} \in \R^{N_\mathrm{coupling}\times N_\mathrm{coupling}}$. For clarity, normalized vector $u/\|u\|_2$ is denoted as $\overline{u}$. The elements of the flux matrix are
\begin{equation}\label{eq:Fab}
    \mathbf{F}_{ab} = \begin{dcases*} \exp\left(-\int_{r_j^i}^{r_l^k}\mu_e(z)\mathrm{d}z\right)p'\left(r_j^i,r_l^k,r_n^m\right), \;\mathrm{if}\; L_a = \left(r_j^i,r_l^k\right) \;\mathrm{and}\; L_b = \left(r_l^k,r_n^m\right),
    \\0, \mathrm{otherwise},
    \end{dcases*}
\end{equation}
where 
\begin{equation}
p'\left(r_j^i,r_l^k,r_n^m\right) = \mu_s\left(r_l^k\right)\int_\Sigma\Theta\left(r_l^k,\overline{(r_l^k - r_j^i)}\cdot \hat{s}'\right)\mathrm{d}\hat{s}'.
\label{eq:F_term_phase}
\end{equation}
The sphere surface $S_2$ subset $\Sigma$ in Eq. \ref{eq:F_term_phase} is defined as the points $\hat{s} \in S_2$ for which $\hat{s}\cdot\overline{(r_n^m - r_l^k)} > \hat{s}\cdot\overline{(r_y^x - r_l^k)} $, $ (r_l^k,r_y^x) \in L$ and $(r_y^x - r_l^k) \neq (r_n^m - r_l^k)$. This means that all the radiation scattered to $\Sigma$ is assumed to scatter toward $r_n^m$, akin to multi-stream approximation. In other words, the scattering phase function at point $r_l^k$ is integrated so that it represents the scattered radiation from the direction of $r_j^i$ towards $r_n^m$, while taking into account the other coupled directions. For example, if one region of $S_2$ has many coupled directions, they get a smaller solid angle space, while another direction with no other couplings nearby could get a whole hemisphere. The coupling towards exactly the same direction is disregarded in this calculation. A simple example on constructing the flux matrix is presented in Sec. \ref{sec:flux_matrix_example}.

The radiation flux $\varphi$ between the nodes can be computed from the system
\begin{equation}
    (I - \mathbf{F})\varphi = S,
\label{eq:flux_system}
\end{equation}
where $I$ is the unit matrix and $S \in \R^{N_\mathrm{coupling}\times 1}$ is the external radiation source contribution, in which the elements
\begin{equation}
    S_a = \exp\left(-\int_{r_s}^{r_j^i}\mu_e(z)\mathrm{d}z\right)p'\left(r_s,r_j^i,r_l^k\right)\phi_0\left(r_s,\overline{(r_j^i - r_s)}\right),
\end{equation}
where $r_s\in\partial\Omega$ is the point on the medium boundary pointing at which $\phi_0(r_s,\overline{(r_j^i - r_s)}) > 0$. Solving the system in Eq. \ref{eq:flux_system} is equivalent to calculating the infinite sum presented in Eq. \ref{eq:fs_sum}. After solving the linear system in Eq. \ref{eq:flux_system}, the $\varphi$ contains the elements
\begin{equation}
    \varphi_a = \phi\left(r_j^i,\overline{(r_l^k - r_j^i)}\right).
\end{equation}

Finally, the radiance observed by the instrument $\phi(r_0^i,-\hat{s}_0^i)$ can be written as the sum
\begin{multline}
    \phi(r_0^i,-\hat{s}_0^i) = \sum_{j=1}^{N_\mathrm{step}^i}\Bigg(\exp\left(-\int_{r_j^i}^{r_0^i}\mu_e(z)\mathrm{d}z\right)\\\cdot\Bigg[\mu_s\left(r_j^i\right)\Theta\left(r_j^i,\overline{(r_{j-1}^i - r_j^i)}\cdot\overline{(r_j^i - r_s)}\right)\exp\left(-\int_{r_s}^{r_j^i}\mu_e(z)\mathrm{d}z\right)\phi_0\left(r_s,\overline{(r_j^i - r_s)}\right) \\+ \sum_{r_l^k\in L'(i,j)}\mu_s\left(r_j^i\right)\Theta\left(r_j^i,\overline{(r_{j-1}^i - r_j^i)}\cdot\overline{(r_l^k - r_j^i)}\right)\phi\left(r_j^i,\overline{(r_l^k - r_j^i)}\right)\Bigg]\Bigg),
\end{multline}
where $L'(i,j) = \{x \;|\; (r_j^i,x) \in L\}$.

\subsection{Simple example of the flux matrix construction}\label{sec:flux_matrix_example}
\begin{figure}
    \centering
    \includegraphics[width=0.5\linewidth]{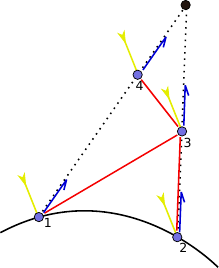}
    \caption{A simple example case with two lines-of-sight (dotted lines) and four nodes (blue circles). The lines-of-sight are drawn from the instrument (black circle) up until the planetary surface (black solid line). The red lines indicate which nodes are coupled. Yellow arrows indicate the incident radiation from an external source and blue arrows the radiation which is observed by the instrument.}
    \label{fig:example-graph}
\end{figure}

To demonstrate the Eqs. \ref{eq:Fab} -- \ref{eq:flux_system} further, a simple example case with four scattering nodes is examined.  A diagram of the case is presented in Fig. \ref{fig:example-graph}. 

With the nomenclature of previous section, these nodes would be $r^0_0$, $r^0_1$, $r^1_0$, $r^1_1$, but for simplicity these nodes are to be denoted as $r_1$, $r_2$, $r_3$ and $r_4$ as per Fig. \ref{fig:example-graph}. The scattering graph is $G(\{r_1,r_2,r_2,r_3\},\{(r_1,r_3),(r_2,r_3),(r_4,r_3),(r_3,r_1),(r_3,r_2),(r_3,r_4)\})$. Then the flux matrix $\mathbf{F}\in\R^{6\times6}$ is of form
\begin{equation}
    \mathbf{F} = \begin{bmatrix}
        0 & 0 & 0 & F_{14} & F_{15} & F_{16} \\
        0 & 0 & 0 & F_{24} & F_{25} & F_{26} \\
        0 & 0 & 0 & F_{34} & F_{35} & F_{36} \\
        F_{41} & 0 & 0 & 0 & 0 & 0 \\
        0 & F_{52} & 0 & 0 & 0 & 0 \\
        0 & 0 & F_{63} & 0 & 0 & 0 \\
    \end{bmatrix}.
\end{equation}
For example, the element $F_{34}$, which describes how much of the radiation originating from $r_4$ is attenuated while propagating to $r_3$ and then scattered at $r_3$ toward $r_1$ is
\begin{equation}\label{eq:example-flux-element}
    F_{34} = \exp\left(-\int_{r_4}^{r_3}\mu_e(z)\mathrm{d}z\right)\mu_s(r_3)\int_\Sigma\Theta\left(r_3,\overline{r_3-r_4}\cdot\hat{s}'\right)\mathrm{d}\hat{s}'.
\end{equation}

%\begin{enumerate}
%      \item The observation lines-of-sight are traced through the medium.
%      \item Depending on the medium properties, \textit{scattering nodes} are created along the line-of-sight during the trace.
%      \item The nodes are paired if \textit{scattering coupling} between them is large enough.
%      \item The Beer-Lambert-Bouguer attenuation along the lines-of-sight and the node pairs are computed.
%      \item With these line attenuations and scattering phase functions, we construct \textit{flux system} with which we can solve observed radiance.
%  \end{enumerate}
%\subsection{Discrete ordinates as a special case}
\section{TURSCA model}\label{sec:tursca}
\begin{figure}
    \centering
    \begin{tabular}{ccc}
       \includegraphics[width=0.3\linewidth]{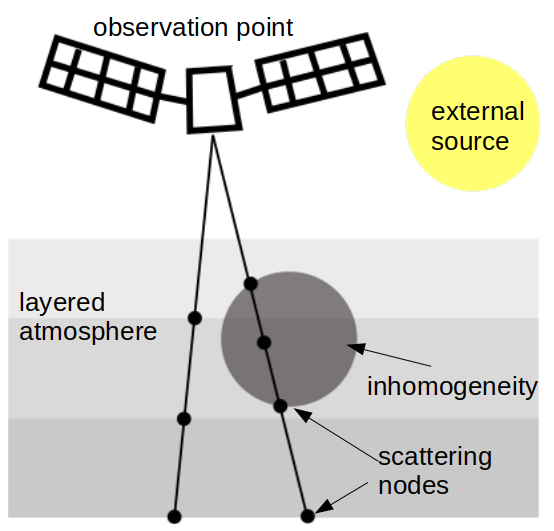}  & \includegraphics[width=0.3\linewidth]{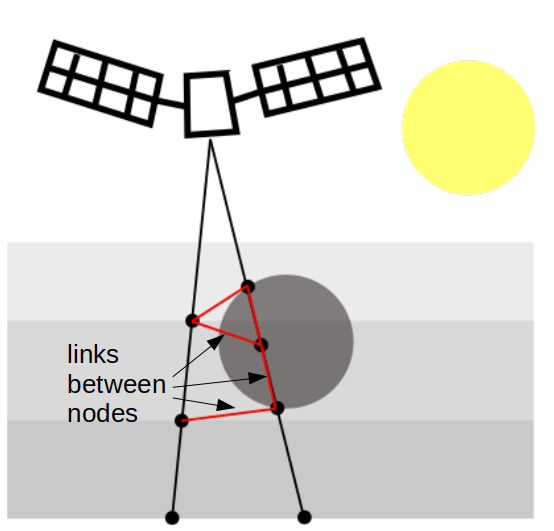} & \includegraphics[width=0.3\linewidth]{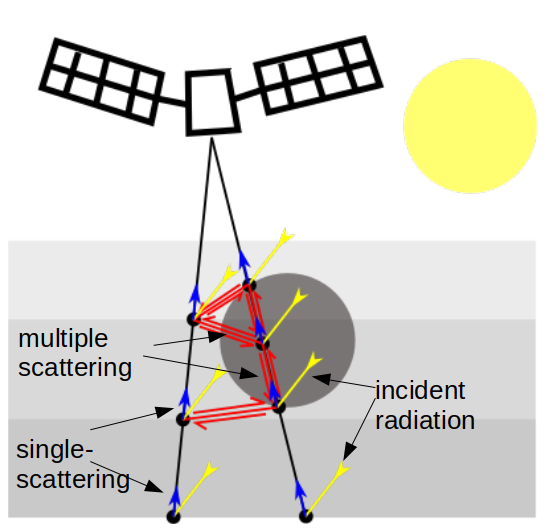} \\
       (a) & (b) & (c)
    \end{tabular}
    \caption{Schematic of the elements found in the TURSCA model. a) The lines-of-sight are traced from the observation point through the atmospheric medium. Depending on the medium basis functions, here the atmospheric layers and a circular inclusion, the scattering nodes are created along the line of sight. b) The scattering nodes are linked, or coupled, if the scattering at the nodes is high enough and they are close enough to each other. c) The incident radiation at each node is used to compute the radiation fluxes between the coupled nodes. The radiance observed by the instrument is then the sum of the single-scattered both incident radiation and the in-between-nodes radiation flux.}
    \label{fig:tursca-comic}
\end{figure}

To compute the results using the method presented in Sec. \ref{sec:theory}, Transmittance calcUlator with Radiative Scattering Coupling Approach (TURSCA) model was created. While the scattering graph method aims to be applicable to general RT problems, TURSCA has been developed with satellite remote sensing in mind. The steps in the computation algorithm are as follows:
\begin{enumerate}
    \item Trace the lines-of-sight from the instrument through the atmosphere and create the scattering nodes along them depending on the medium basis functions.
    \item Compute the scattering coupling between all the node pairs and couple them if the coupling value is higher than a preset threshold and there are less total couplings than a preset maximum amount. If there would be more couplings than maximum amount, then only the couplings with the highest coupling value are selected.
    \item Construct the matrix $\mathbf{F}$ and the vector $S$ by evaluating the phase function integrals and Beer-Lambert-Bouguer attenuation between the scattering nodes and the radiation source.
    \item Solve the constructed flux system (Eq. \ref{eq:flux_system}) to obtain the radiance field vector $\varphi$.
    \item Compute the single-scattered radiance toward the instrument at each of the scattering nodes from $\varphi$ and the external source, and attenuate it according to Beer-Lambert-Bouguer on its path to the instrument.
\end{enumerate}
\begin{figure}
    \centering
    \includegraphics[width=\linewidth]{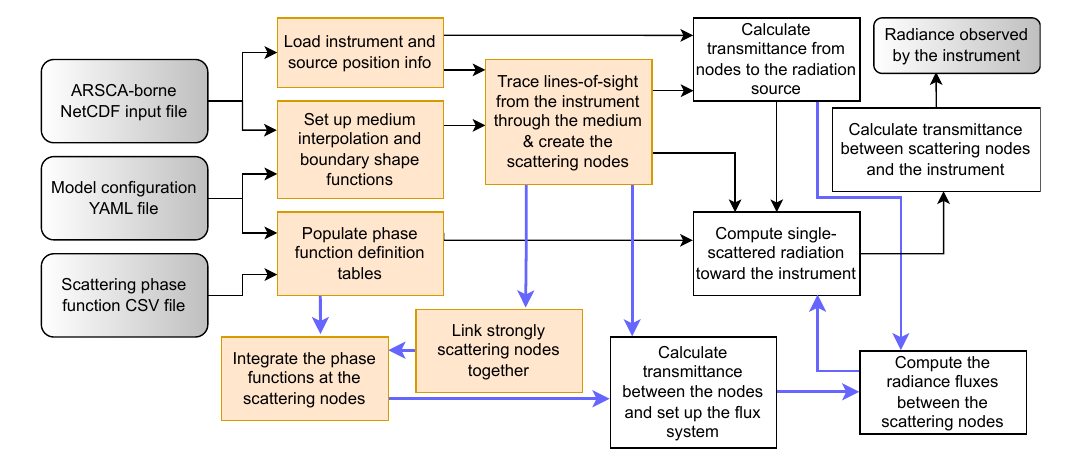}
    \caption{Diagram visualizing the computation flow within TURSCA. The rounded rectangles with silver background represent input and output files. Rectangles represent different steps of the algorithm. TURSCA may be used to compute single-scattered radiances by following only the black arrows and multiple scattered radiances by also following the blue arrows. If scattering is assumed to be constant within a wavelength band and only the absorptivity varies, the rectangles with orange tint need to be computed only once and may be reused for subsequent wavelengths.}
    \label{fig:tursca-data}
\end{figure}
A data flow diagram within TURSCA is presented in Fig. \ref{fig:tursca-data}. A visualization of the RT computation within TURSCA is presented in Fig. \ref{fig:tursca-comic}. The rest of this section describes the technical details of the algorithm implementation.

The code is publicly available on Github (\url{https://github.com/amikko/tursca}) and it may be used freely in accordance to the MIT License. TURSCA is run on the command line using Python 3 and it requires NumPy, SciPy, NetCDF and Taichi libraries. The input parameters are defined in a NetCDF file created with Atmospheric Radiation Simulation Computation Application (ARSCA) \cite{MIKKONEN2024108892}, which is also available on Github (\url{https://github.com/amikko/arsca}). Model configuration are defined in YAML-formatted text file.

%\begin{itemize}
%    \item {\tiny configs}
%    \item future development!
%    \item {\tiny line integration, basis contrib.}
%    \item {\tiny phase function integration, fibodirs}
%    \item {\tiny link PCA -muunnos}
%    \item {\tiny steps, basis functions}
%    \item {\tiny boundaries, light sources}
%    \item {\tiny taichi: CPU and GPU}
%    \item usage and installation, github-linkki!
%    \item käy koodi lävitte ja kirjaa ylös erikoisuudet!
%    \item näistä sitten koonti tuohon ekaan paragrafiin!
%\end{itemize}

The medium properties in TURSCA are defined in three different ways: flat slabs, spherical shells or 3D Gaussian clouds. These definitions are presented in the form of medium basis functions from which the medium properties can be calculated at arbitrary medium points. The flat slab and spherical shell basis functions have definite values at predefined altitudes and the values between these altitudes are linearly interpolated from the closest points. The Gaussian cloud has a position and a standard deviation and its contribution is summed together with other basis functions. The Gaussian cloud is assumed to have no effect at a distance more than two standard deviations from its center. TURSCA also creates the scattering nodes in accordance to these basis functions. When tracing a line-of-sight through the medium, scattering nodes are created at slab and shell definition altitudes, or dotted at previously defined minimum step length when closer than two standard deviations to a Gaussian cloud center.

The medium boundary geometries can be parallel planes or concentric sphere surfaces. For radiative transfer, surfaces can be defined as Lambertian reflectors, or pass-through, which does not block radiation. The Lambertian surfaces can cast shadows, which becomes apparent with spherical surfaces: scattering node at the surface is not illuminated by the radiation source if the local solar zenith angle is over $\pi/2$, or by radiation originating from another scattering node if the line segment connecting the nodes is not fully within the medium. For scattering coupling calculations (Eq. \ref{eq:scatt_coup}), the nodes at a Lambertian surface have a $\mu_s^\mathrm{Lambertian} = A\cdot\max\{\mu_s(r)\}$, where $A$ is the Lambertian albedo. Currently the only modelled radiation source is a far-field source originating from outside the medium. Both surface geometry and reflectance as well as radiation source modalities are easily added to TURSCA if desired.

TURSCA requires the computation of several different path integrals, such as from the illumination source to scattering nodes, between two scattering nodes and from a scattering node to the instrument. These paths are stored as arrays denoting how much of the path is situated in each of the medium basis functions. This is useful when integrals over different quantities, such as extinction or scattering at different wavelengths, is needed.

Compared to the scattering graph construction presented in Sec. \ref{sec:theory}, TURSCA employs a different, but equivalent, strategy. Instead of selecting a particular value for $c_\mathrm{min}$, the maximum amount of couplings is preselected before running the model. This is more convenient from the computational point-of-view, because an adequate amount of memory can be reserved for $\mathbf{F}$, for example. The scattering node couplings are done by starting with the highest coupling value between the nodes and counting down from there. Thus, the smallest coupling value included in the coupling array is effectively the $c_\mathrm{min}$.

For scattering coupling, the relative positions rather than absolute positions of the scattering nodes are of importance. Because of this, for the purposes of scattering coupling computations, the absolute scattering node positions are zero-meaned, their standard deviation is normalized to one in each direction and they are mapped onto their eigenvectors, i.e. principal components, transforming them into relative node positions. With this approach the differences in spatial resolutions and viewing angles between different scenes do not result into fundamentally different coupling behaviour.

\begin{figure}
    \centering
    \includegraphics[width=0.45\linewidth]{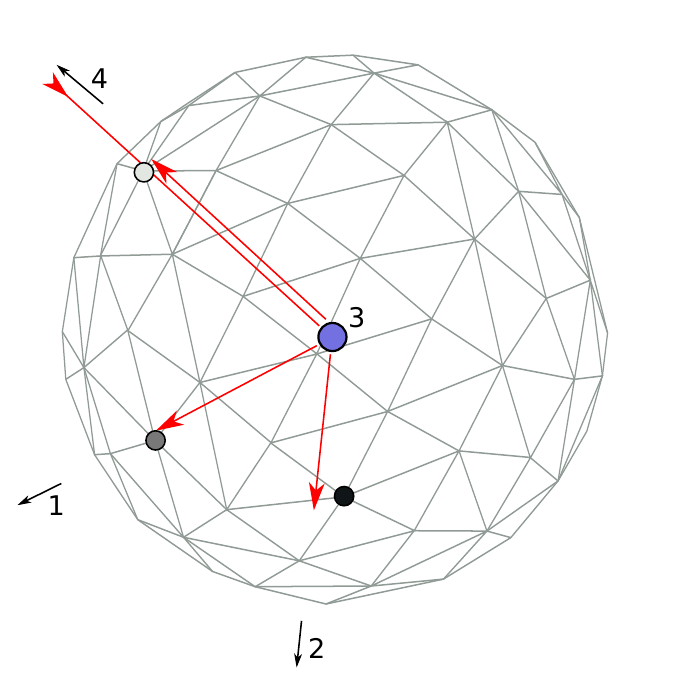}
    \includegraphics[width=0.45\linewidth]{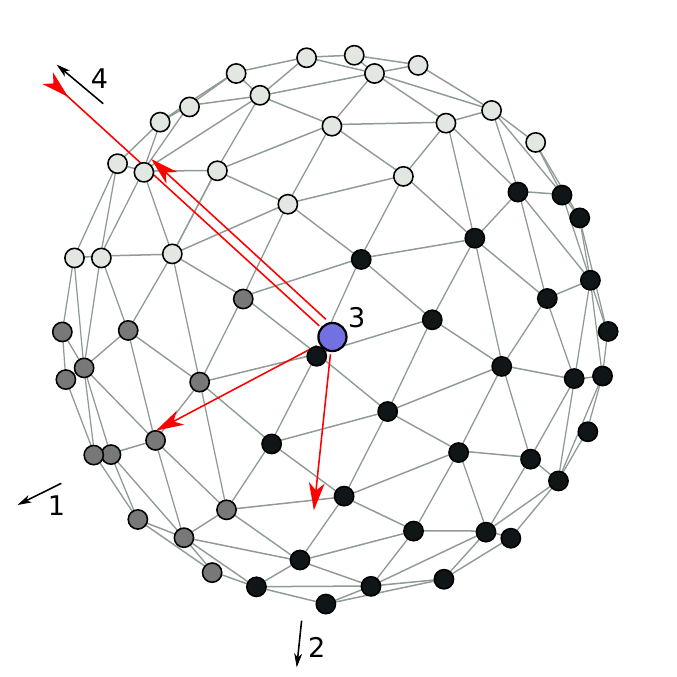}
    \caption{Demonstration of the flood fill in one hemisphere of the Fibonacci lattice of 100 points at $r_3$ scattering the radiation from $r_4$ in the example case presented in Sec. \ref{sec:flux_matrix_example}. In the left image, the closest lattice point toward each scattering direction is assigned a color: gray for $r_1$, black for $r_2$, and white for $r_4$. In the right image, the whole lattice if filled with the three colors.}
    \label{fig:fibonacci-integral-example}
\end{figure}

TURSCA can simulate Rayleigh scattering as well scattering from randomly-oriented particles with an arbitrary phase function defined in the form of a look-up table. The scattering phase functions need to be integrated as described in Eq. \ref{eq:F_term_phase}. This is done by using a Fibonacci lattice \cite{gonzalez2010measurement}. For each medium basis function, each scattering phase function is evaluated toward each Fibonacci lattice point. As the Fibonacci lattice is generated approximately evenly around the sphere surface, highly asymmetric phase functions may require large number of lattice points to be accurately represented. The resulting scattering phase function value table is then used to compute the $p'$ so that each $\overline{(r_n^m - r_l^k)}$, for some $r_j^i$ and $r_l^k$, is rotated so that the forward scattering direction $\theta = 0$ coincides with the $\theta = 0$ direction of the table. The rotated $\overline{(r_n^m - r_l^k)}$ are then assigned to the closest lattice point, which is then colored (i.e. assigned an integer) with a color corresponding to that particular $\overline{(r_n^m - r_l^k)}$. Then the mostly uncolored lattice is then colored fully using a flood fill method, where neighbouring uncolored lattice points are colored in iteratively with their neighbours' colors. An example of this process is presented in Fig. \ref{fig:fibonacci-integral-example}. The adjacency information of the Fibonacci lattice points is obtained by the edges of polyhedron defined by the convex hull of the lattice point set. The integral of the phase function over $\Sigma$ defined by $\overline{(r_n^m - r_l^k)}$ is then calculated as a sum of phase function values at lattice points with that particular color. 

Commonly in discrete ordinates approaches a level symmetric even quadrature set is used (e.g. \cite{manalo2015advanced}). In the case of scattering graph method such approach is not possible due to the inherent irregularity of the scattering directions between the nodes. The Fibonacci sphere approach simply distributes the directions onto the unit sphere approximately evenly. By having 100 Fibonacci directions, for example, the sphere is divided into cones with solid angle of $4\pi/100$. The cones have an apex angle of about 22 degrees, which can be interpreted as the scattering phase function angle discretization.

%\begin{figure}
%    \centering
%    \includegraphics[width=\textwidth]{tursca-2.png}
%    \caption{TURSCA. Credit: Tomi Karppinen}
%    \label{fig:enter-label} 
%\end{figure}
%\subsection{Performance tests of each step of the algorithm}
%\subsection{Performance on CPU and LUMI supercomputer}

The technical implementation of TURSCA was done using Taichi, a domain-specific programming language for parallel computing embedded in Python \cite{taichi-2019}. Taichi enables the development of back-end agnostic code in pure Python syntax, which means that same program can be run on CPUs and GPUs. The language features support the creation of sparse data structures, which are used extensively in TURSCA, and high level of control on their access patterns, which allow further technical optimization of the TURSCA code, and automatic differentiation for future retrieval algorithm development. TURSCA code has been tested on an Asus GeForce RTX 3090 GPU as well as several different CPUs.

%Future development of TURSCA includes the computation of full Stokes polarization vector and thermal emissivity of the medium, which both require further validation studies. The bending of the optical path due to atmospheric refraction could be implemented for line-of-sight tracing. Spatially varying non-Lambertian surface bi-directional reflectance distribution functions as well as different types of radiation sources and medium basis functions can be implemented with relative ease. Optimization for computation speed and memory usage are in the works, especially regarding RT simulations spanning several wavelengths.

%{\color{blue} TODO: tehtävä vuokaavio kulusta!}
\section{Validation against other 3D RT codes}
To examine the validity of the TURSCA model, comparisons against established RT solvers were carried out. Two different RT solvers, Siro and SHDOM, were selected to get a range of features examined and they were run with the same input parameters as TURSCA. These comparison models were selected because of their different approaches to RT modeling and different capabilities of 3D RT, demonstrating the generality of TURSCA. 

Siro is a backward Monte Carlo RT model first presented by \cite{oikarinen1999multiple} to simulate UV-NIR atmospheric observations by limb-viewing satellites in atmospheres defined as spherical shells. In Siro, different orders of scattering can be examined separately, which enriches its capacity for analyzing scattering phenomena. Recently Siro has been used as a reference model in RT model intercomparison, where it agreed within 1--3\% with the other models in limb-viewing geometries \cite{zawada2021systematic}. 

SHDOM is a spherical harmonic and discrete ordinates method based model for RT in 1-, 2- and 3-D atmospheres, presented initially by \cite{shdom}. Recently it has been used for cloud tomography \cite{rs12172831}, analysis of rocket plumes \cite{ZHANG2022104054} and 3D effects in satellite remote sensing of carbon dioxide \cite{amt-16-2145-2023}. Since its inception, it has been extended in various ways, such as spectral RT \cite{DOICU2021107386}. In a comprehensive study of model performance, SHDOM was shown to agree with the I3RC Monte Carlo model within 1\% in nadir-viewing cloudy 3D scenes.

\subsection{Validation against Siro}
Siro and TURSCA RT simulations of top-of-the-atmosphere transmittances at the wavelength of 0.765 $\mu$m were carried out in a spherical atmosphere. Specifics of the simulation are presented in the Table \ref{tab:siro_simu_specs}. Siro was run simulating 1 million photons with 0.1 km step length. TURSCA was run with 100 Fibonacci directions and 2700 scattering couplings in the clear sky case and 5700 scattering couplings in the aerosol case. Additionally, TURSCA was simulated with 4 line-of-sight configurations: 1$\times$1, 3$\times$3, 5$\times$5 and 7$\times$7. The center pixel is exactly at the VZA, but for accurate modeling of the scattering, TURSCA requires several lines-of-sight, which are cast from the instrument position at angles {0.494$^\circ$$\;\times\;$0.494$^\circ$}, {0.824$^\circ$$\;\times\;$0.824$^\circ$} and {1.154$^\circ$$\;\times\;$1.154$^\circ$}, respectively.

\begin{figure}
    \centering
    \includegraphics[width=\linewidth]{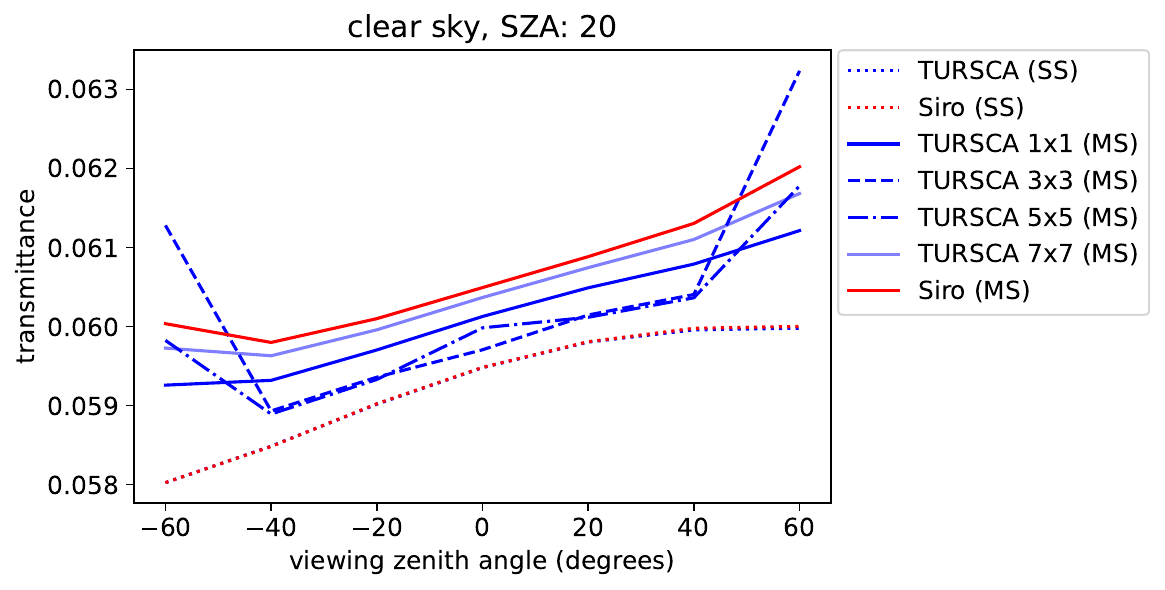}
    \includegraphics[width=\linewidth]{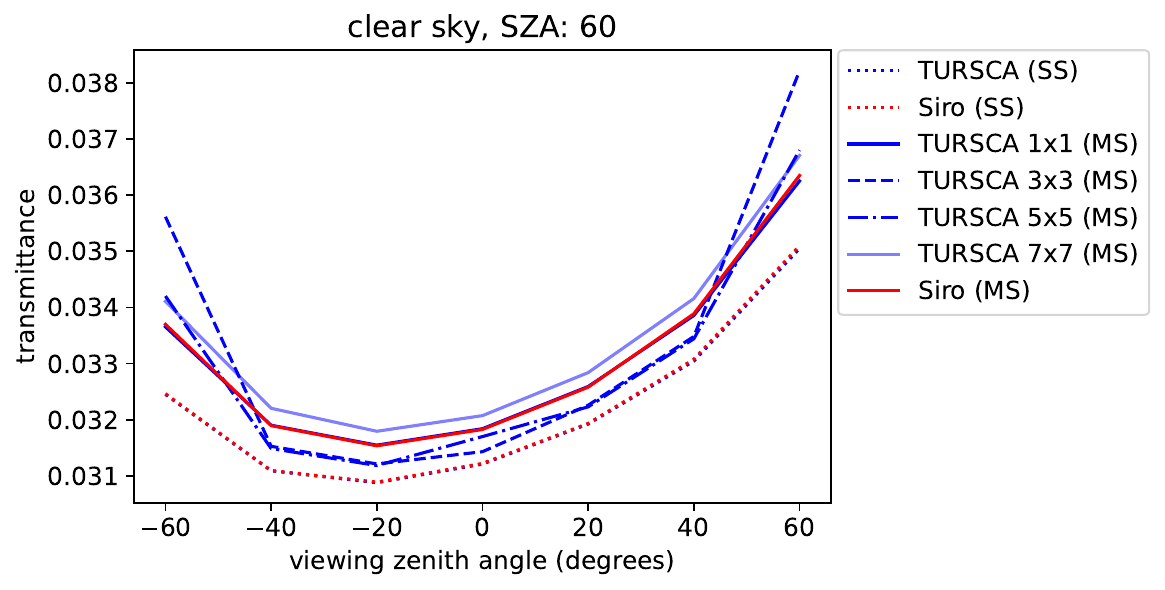}
    \caption{Transmittances simulated by Siro and TURSCA in the clear sky case. The dotted lines are the single-scattering (SS) and other styles are multiple scattering (MS). For TURSCA results, only the transmittance of the central pixel is presented in 3$\times$3, 5$\times$5 and 7$\times$7 cases.}
    \label{fig:siro_valid_clear}
\end{figure}

\begin{figure}
    \centering
    \includegraphics[width=\linewidth]{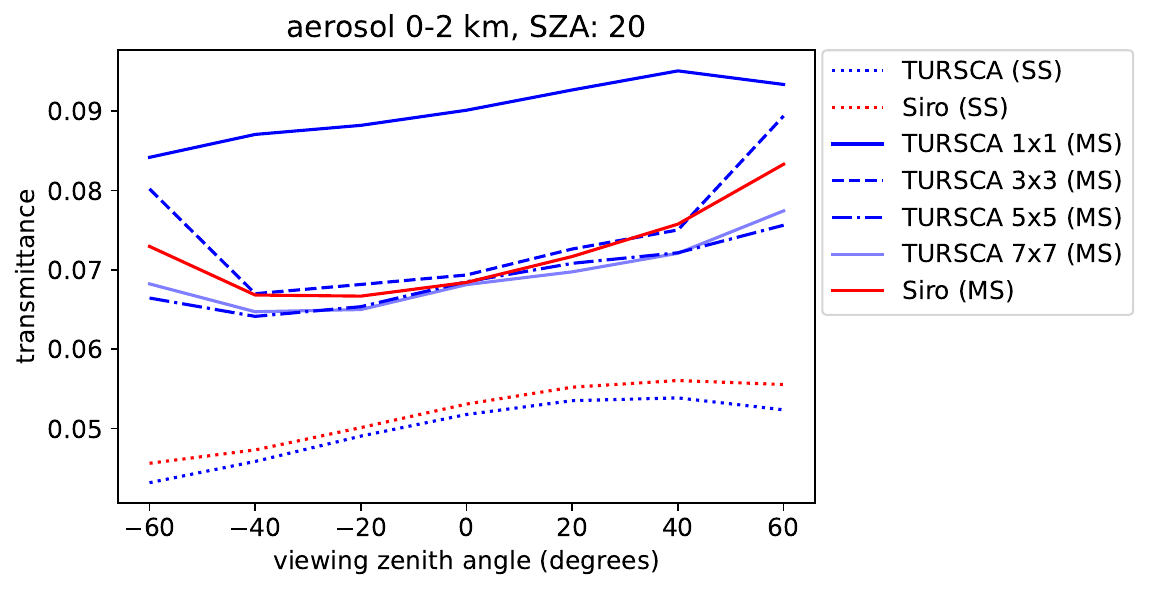}
    \includegraphics[width=\linewidth]{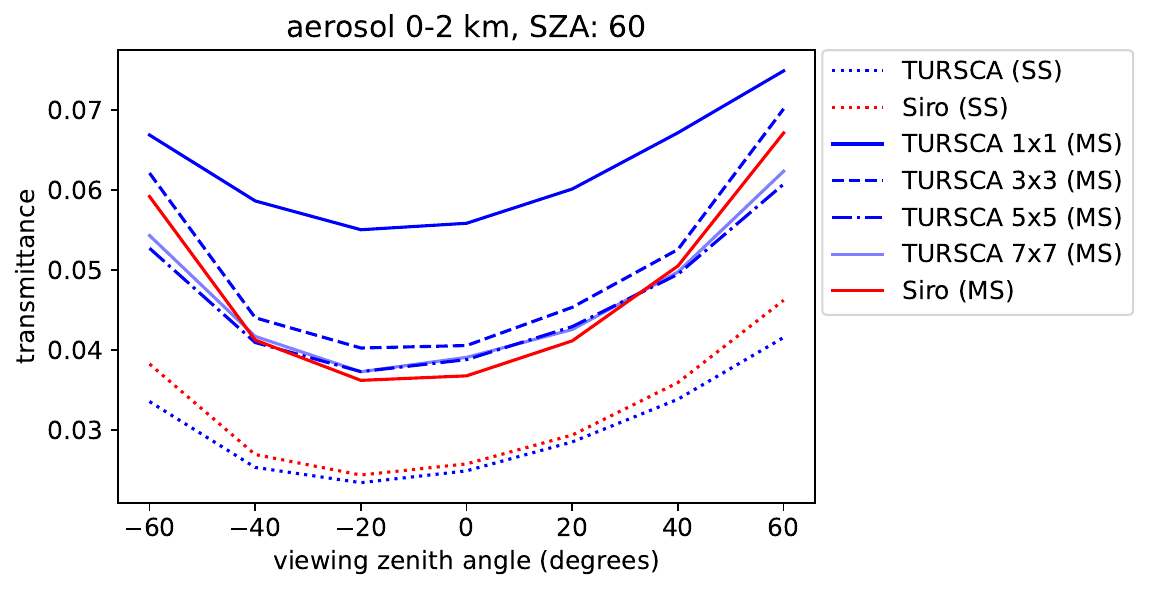}
    \caption{Transmittances simulated by Siro and TURSCA in the aerosol case. The dotted lines are the single-scattering (SS) and other styles are multiple scattering (MS). For TURSCA results, only the transmittance of the central pixel is presented in 3$\times$3, 5$\times$5 and 7$\times$7 cases.}
    \label{fig:siro_valid_aerosol}
\end{figure}

\begin{table}[]
    \centering
    \caption{Specifics of Siro and TURSCA simulation parameters for the validation simulations.}
    \begin{tabular}{cc}
       \hline Input parameter & Value \\
       \hline Solar zenith angle & \{20,60\} degrees \\
       Viewing zenith angle & \{-60,-40,-20,0,20,40,60\} degrees \\
       Wavelength & 0.765 µm \\
       Lambertian surface albedo & 0.2 \\
       Surface pressure & 1.0139 atm \\
       Instrument altitude & 417 km \\
       Atmosphere & \{clear, constant aerosol layer 0 - 2 km\}\\
       Absorbing gases & None \\
       Aerosol radius & 0.07 $\mu$m\\
       Aerosol refractive index & $1.4 - 0.003i$ \\
       Avg. cosine of phase function & 0.06\\
       Aerosol optical depth & 0.3 \\
       \hline
    \end{tabular}
    
    \label{tab:siro_simu_specs}
\end{table}
The simulation results are presented in Figures \ref{fig:siro_valid_clear} and \ref{fig:siro_valid_aerosol}, and the relative errors are presented in Table \ref{tab:tursca-siro-valid}. The single-scattering results agree almost perfectly in the clear sky case and quite well in the presence of aerosols. The multiple scattering is captured really well in both 1x1 and 7x7 sizes in the clear sky case, which is apparent from TURSCA curve shapes matching Siro in Fig. \ref{fig:siro_valid_clear} and miniscule standard deviation of relative error in Table \ref{tab:tursca-siro-valid}. The best match to Siro curve shapes in Fig. \ref{fig:siro_valid_aerosol} are obtained with size 7x7 for SZA 20 and size 5x5 for SZA 60, which have 2.19\% and 2.84\% standard deviation of relative error, respectively. Low standard deviation of relative error indicates likeness in the modeling of the scattering. 

When considering multiple scattering, the necessity of simulating several lines-of-sight becomes apparent: larger image size tends to lead to more accurate results. In the aerosol case, the 1x1 performs poorly, which is to be expected, but increasing the image size even to 3x3 increases the accuracy significantly. Generally it can be said that for VZAs close to 0, the radiances are quite accurately simulated even with a small image sizes, but for larger VZAs, more pixels are needed. This is an expected result, because increased atmospheric scattering in the presence of aerosols and at larger VZAs requires several lines-of-sight to be accurately modelled.

The differences in the simulation results between Siro and TURSCA stem from the procedure to create the optical paths for the multiply scattered radiation. In Siro, the paths are constructed naturally as a result of the scattering processes themselves, whereas in TURSCA the optical paths are constructed systematically with lines-of-sight and scattering coupling function. This choice of coupling function (Eq. \ref{eq:scatt_coup}) appears to result into similar scattering behaviour between Siro and TURSCA with adequate accuracy.

To gauge the computational performance of these compared RT models, some mean runtimes are presented. For example, in the clear sky case at SZA 20, the mean runtime for a single viewing angle was 34.5 s for 5x5 multiple scattering TURSCA and 314.9 s for Siro. For the aerosol case at SZA 60 the runtimes were 55.9 s and 320.2 s respectively.

To assess the similarity between Siro and TURSCA, the mean of the standard deviations of the relative errors from the Table \ref{tab:tursca-siro-valid}, disregarding the single-scatter and 1x1 multiple scatter, was calculated. With that it can be then said that TURSCA agrees with Siro within 2.47\% $\approx 3$\% on average.


\begin{table}[]
    \centering
    \caption{Relative errors $(T_\mathrm{TURSCA} - T_\mathrm{Siro})/T_\mathrm{Siro}$ and their standard deviations along VZAs from Figs \ref{fig:siro_valid_clear} and \ref{fig:siro_valid_aerosol}.}
    \begin{tabular}{ccccc} \hline
        &SZA 20, clear (\%)&SZA 60, clear (\%)&SZA 20, aerosol (\%)&SZA 60, aerosol (\%)\\ \hline
        single-scatter &  $-0.01 \pm 0.02 $&$-0.06 \pm 0.04$&$-3.68 \pm 1.30$& $-6.33 \pm 3.29$\\
        1x1 mult. scatter &$-0.88 \pm 0.28$&$-0.05 \pm 0.11$&$25.22 \pm 7.60$& $35.65 \pm 15.95$\\
        3x3 mult. scatter &$-1.02 \pm 0.46$&$-0.35 \pm 1.12$&$-4.29 \pm 3.39$& $-1.49 \pm 6.07$\\
        5x5 mult. scatter &$-0.38 \pm 1.51$&$0.74 \pm 2.95$&$3.06 \pm 3.70$&$7.41 \pm 2.84$ \\
        7x7 mult. scatter &$-0.34 \pm 0.13$&$0.91 \pm 0.15$&$-3.88 \pm 2.19$&$-0.43 \pm 5.08$ \\ \hline
    \end{tabular}
    \label{tab:tursca-siro-valid}
\end{table}
\subsection{Validation against SHDOM}
\begin{figure}
    \centering
    \includegraphics[width=0.9\linewidth]{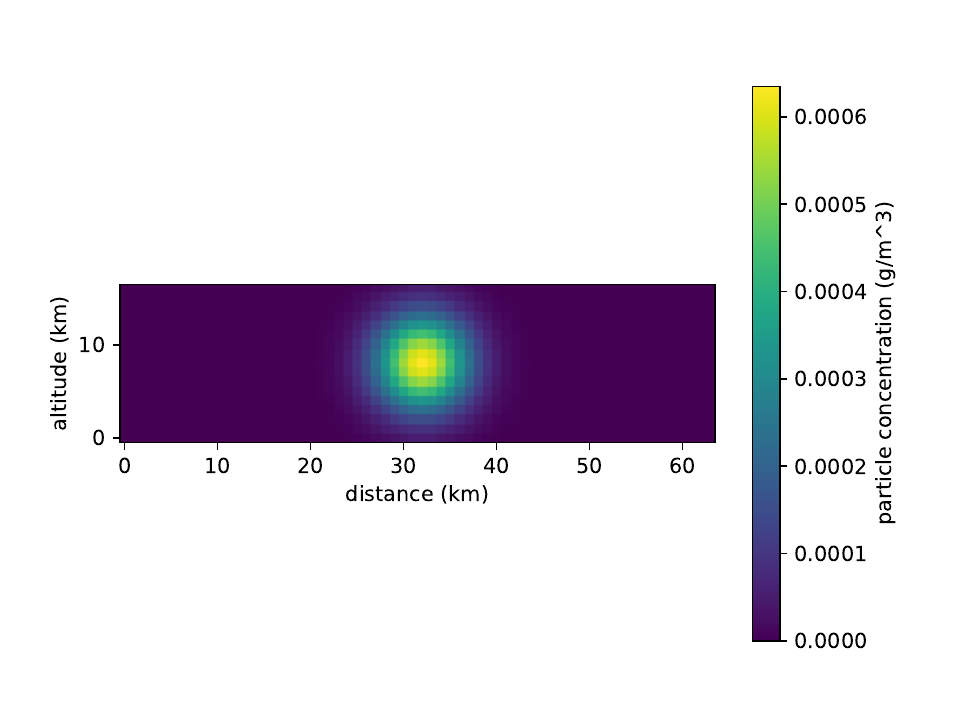}
    \caption{The spatial distribution of cloud density used in SHDOM and TURSCA transmittance simulations.}
    \label{fig:shdom-cloud}
\end{figure}

\begin{figure}
    \centering
    \begin{tabular}{cc}
    \includegraphics[width=0.5\textwidth]{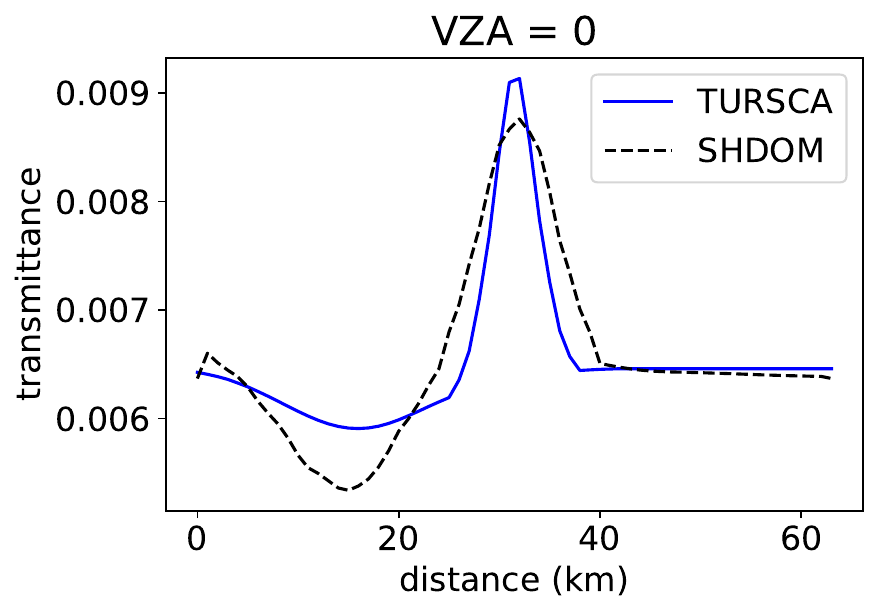}& \includegraphics[width=0.5\textwidth]{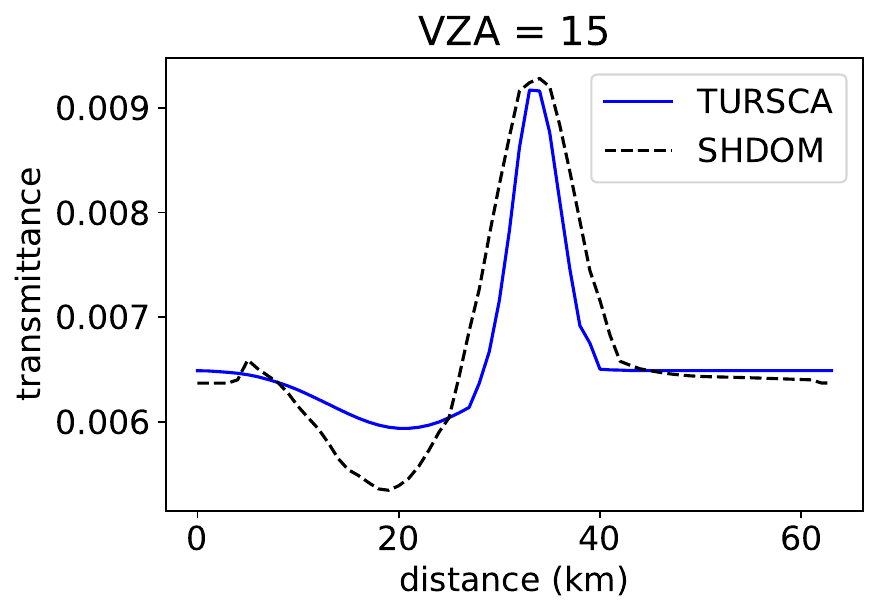} \\
    \includegraphics[width=0.5\textwidth]{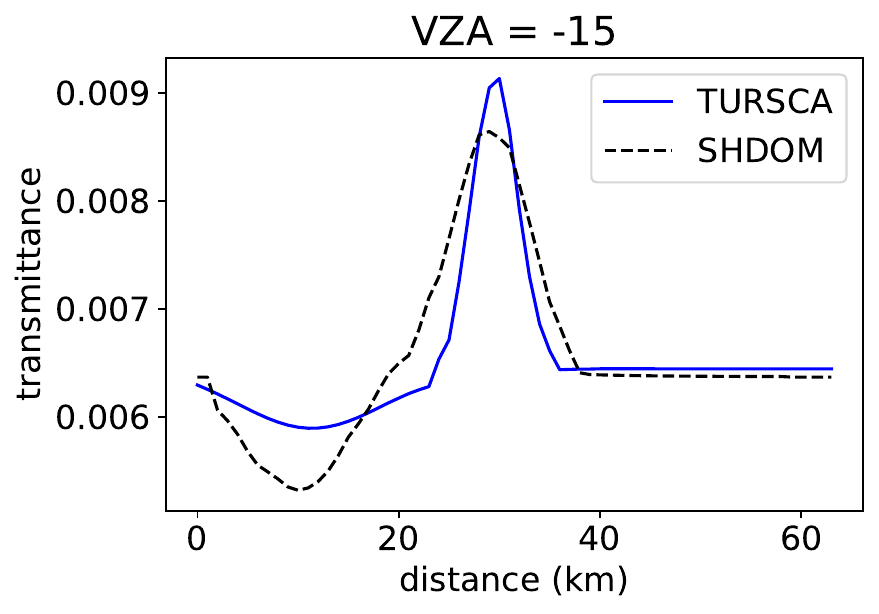}& \includegraphics[width=0.5\textwidth]{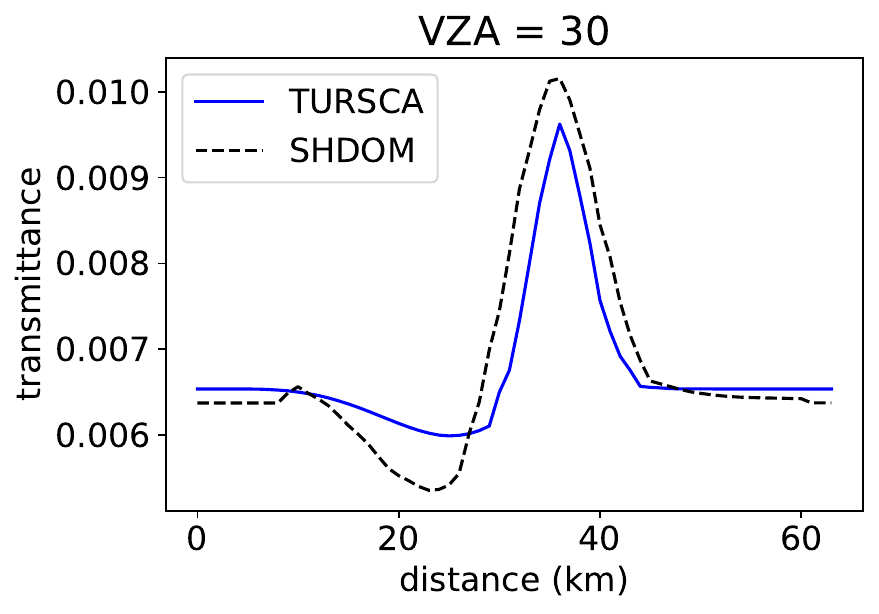}\\
    \includegraphics[width=0.5\textwidth]{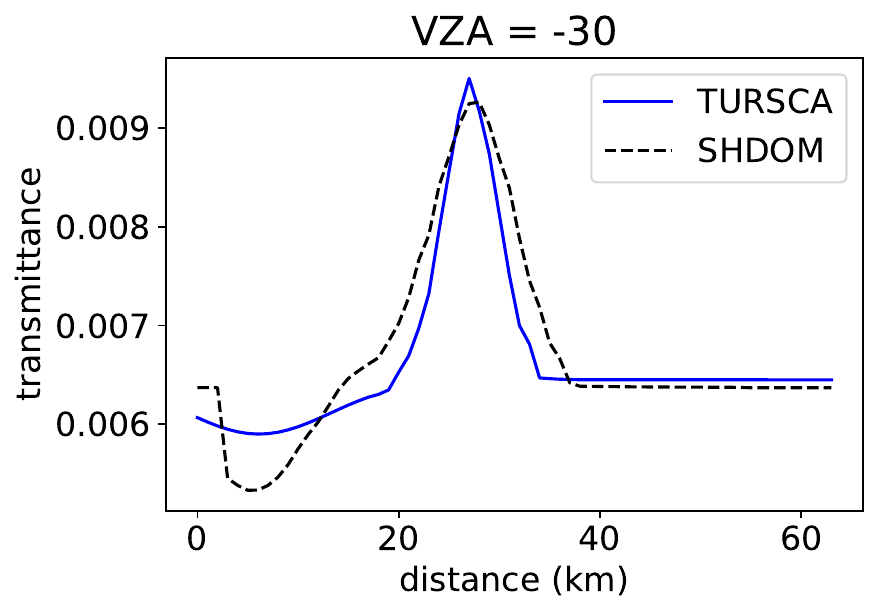}& \includegraphics[width=0.5\textwidth]{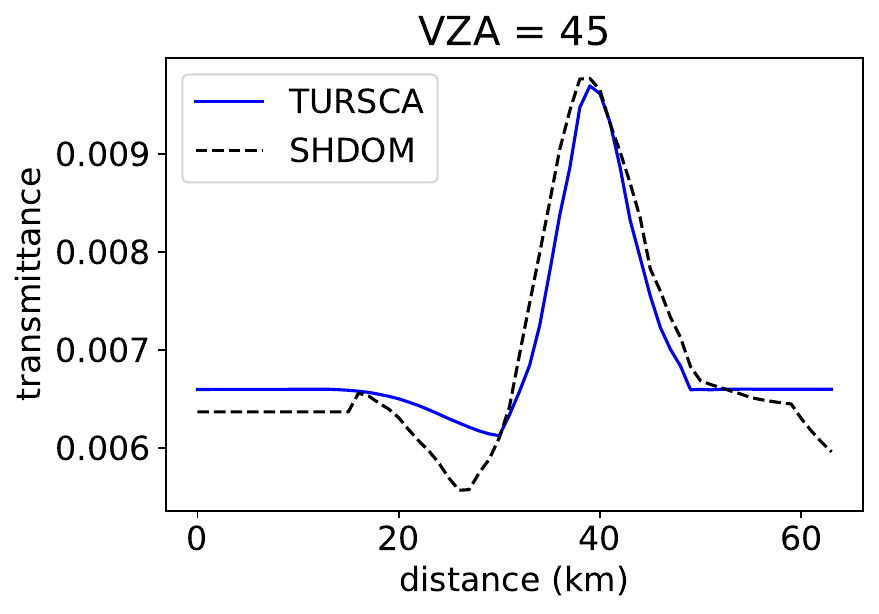} \\
    \includegraphics[width=0.5\textwidth]{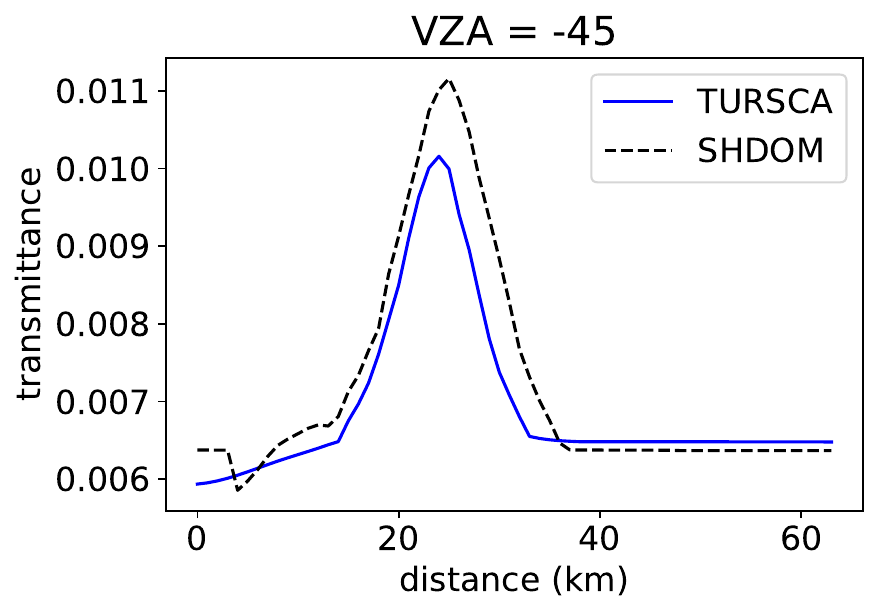}& \includegraphics[width=0.5\textwidth]{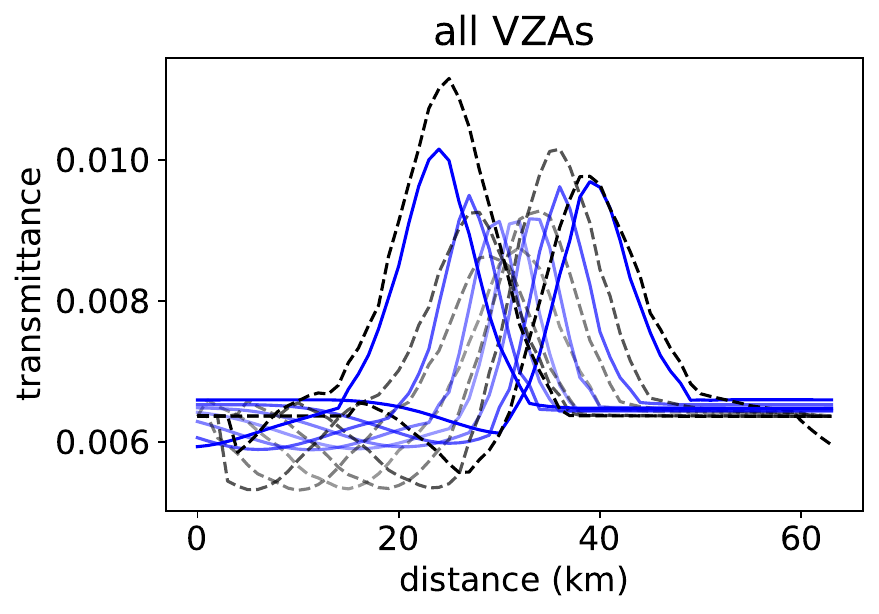}
    \end{tabular}
    \caption{TURSCA and SHDOM top-of-the-atmosphere transmittances at 7 different VZAs simulated according to specifications in Table \ref{tab:shdom_simu_specs} and cloud geometry in Fig. \ref{fig:shdom-cloud} across the 64 km scene with a spherical cloud at 32 km mark. On the bottom right, all 7 transmittances are overlaid.}
    \label{fig:shdom-tursca-validation}
\end{figure}

TURSCA was also validated against SHDOM. A cloudy scene at 1.65 $\mu$m in a 2D atmosphere was simulated with both models to examine the capabilities of TURSCA in higher atmospheric dimensions. The top-of-the-atmosphere transmittances in a 17 km $\times$ 64 km scene containing a single Gaussian water cloud (Fig. \ref{fig:shdom-cloud}) were simulated with both SHDOM and TURSCA. The specifics on the simulation parameters are presented in Tab. \ref{tab:shdom_simu_specs}. SHDOM was run with zero constant horizontal boundary conditions and 32 zenith and azimuth angles. TURSCA was run with 100 Fibonacci directions, 0.5 km minimum step length and 20000 scattering couplings. The simulated transmittances are presented in Fig. \ref{fig:shdom-tursca-validation} and the relative differences between the models are presented in Table \ref{tab:shdom-relerr}.
\begin{table}[]
    \centering
    \caption{Specifics of SHDOM and TURSCA simulation parameters for the validation simulations.}
    \begin{tabular}{cc}
        \hline Input parameter & Value \\
       \hline Solar zenith angle & 60 degrees \\
       Viewing zenith angle & \{-45,-30,-15,0,15,30,45\} degrees \\
       Wavelength & 1.65 µm\\
       Lambertian surface albedo & 0.04 \\
       Surface pressure & 0.999 atm \\
       Atmosphere & water cloud, shown in Fig. \ref{fig:shdom-cloud}\\
       Absorbing gases & None \\
       Aerosol radius & 2.0 $\mu$m\\
       Aerosol refractive index & $1.4 - 0.003i$ \\
       Avg. cosine of phase function & 0.84\\
       \hline
    \end{tabular}
    \label{tab:shdom_simu_specs}
\end{table}

\begin{table}[]
    \centering
    \caption{Mean and standard deviation of relative error $(T_\mathrm{TURSCA} - T_\mathrm{SHDOM})/T_\mathrm{SHDOM}$ along the 64 lines-of-sight for each of the viewing zenith angles (VZA).}
    \begin{tabular}{cc} \hline
        VZA (degrees) & Relative error (\%) \\ \hline
        0 & $0.15 \pm 5.30$ \\
        15 & $-0.03 \pm 6.12$\\
        30 & $-0.10 \pm 7.01$ \\
        45 & $1.48 \pm 4.85$\\
        -15 & $0.51 \pm 5.28$\\
        -30 & $-0.61 \pm 5.15$\\
        -45 & $-3.05 \pm 5.50$\\ \hline
    \end{tabular}
    
    \label{tab:shdom-relerr}
\end{table}
Qualitatively examining Fig. \ref{fig:shdom-tursca-validation}, the location of the transmittance peaks, i.e. the maximum of cloud scattered radiation matches well between the models. However difference in the height of the peaks varies by the VZA, which could be an effect of the highly forward-scattering phase function ($g = 0.84$) of the cloud particles, because of the isotropic nature of scattering coupling in TURSCA. The most forward-scattering direction (VZA = -45 degrees) exhibits the largest difference between the models, which could be indicative of this. Additionally, the anisotropy effects can be seen from the bottom right image in the figure, where all the transmittances are plotted together: TURSCA peak heights are distributed relatively symmetrically around the 32 km distance mark, whereas SHDOM peaks heights are asymmetrically distributed, as is intuitively expected.

Another culprit for the peak height differences could be the longer range scattering effects between the surface and the cloud not fully captured by TURSCA due to its distance-dependent coupling function. This could also be the reason for the deeper shadow cast by the cloud in SHDOM and slightly different position of the shadow transmittance minimum. Sharp indentations in the SHDOM transmittance near the edges are most likely caused by the boundary conditions.

The very faint decline in the SHDOM transmittance from 40 to 60 km visible in the VZAs shown in the left column of Fig. \ref{fig:shdom-tursca-validation} is probably the result of scattering between the cloud particles and the Rayleigh scattering of the air molecules. In TURSCA, this line is flat because the scattering coupling is quite low due to the low Rayleigh scattering cross-sections in 1.65 $\mu m$ wavelength.

The standard deviation of relative errors presented in Table \ref{tab:shdom-relerr} are fairly large, roughly about 6\%, but the mean of the relative errors is less than 1\%, barring the VZA 45$^\circ$ and -45$^\circ$ cases. This indicates that the total observed radiance at the top-of-the-atmosphere agrees well between the models, but its spatial distribution has discrepancies. These differences could arise from differing spatial and angular discretizations between the cells in SHDOM and scattering nodes in TURSCA.

Based on this comparison, it can be concluded that TURSCA can simulate top-of-the-atmosphere transmittances in 2D atmospheres with a satisfactory accuracy. However, when simulating highly anisotropic scattering phase functions, TURSCA could benefit from a different definition of Eq. \ref{eq:scatt_coup}, possibly tied to the asymmetry parameter of the phase functions.

To gauge the computational performance of these models, the runtime for SHDOM was 1.1s for the whole dataset whereas TURSCA on average took 207.4 s per viewing angle with the total of 1244.1 s for whole runtime.

To assess the similarity between SHDOM and TURSCA, the mean of the standard deviations of the relative errors from the Table \ref{tab:shdom-relerr}, alike in the previous section, was calculated. With that it can be then said that TURSCA agrees with SHDOM within 5.60\% $\approx 6$\% on average.

\section{The transmittance as a function of the amount of scattering coupling}\label{sec:coupling_effect}
\begin{figure}
    \centering
    \includegraphics[width=1.0\linewidth]{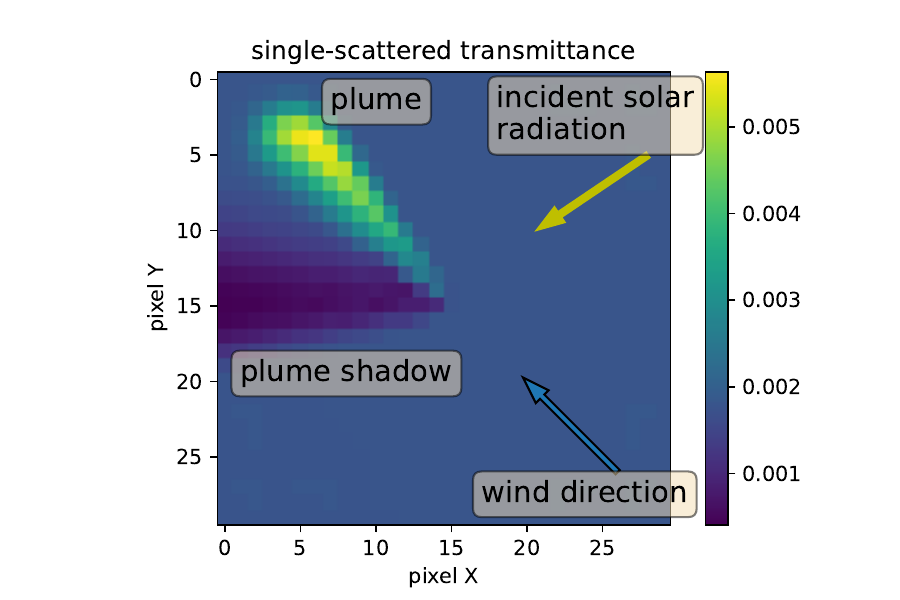}
    \caption{A 30$\times$30 pixel image of the top-of-the-atmosphere single-scattered transmittance in the plume scene examined in Sec. \ref{sec:coupling_effect}. The wind-carried plume scatters radiation as well as casts a shadow.}
    \label{fig:plume_ss_30}
\end{figure}

\begin{figure}[h]
    \centering   
    \includegraphics[width=1\textwidth]{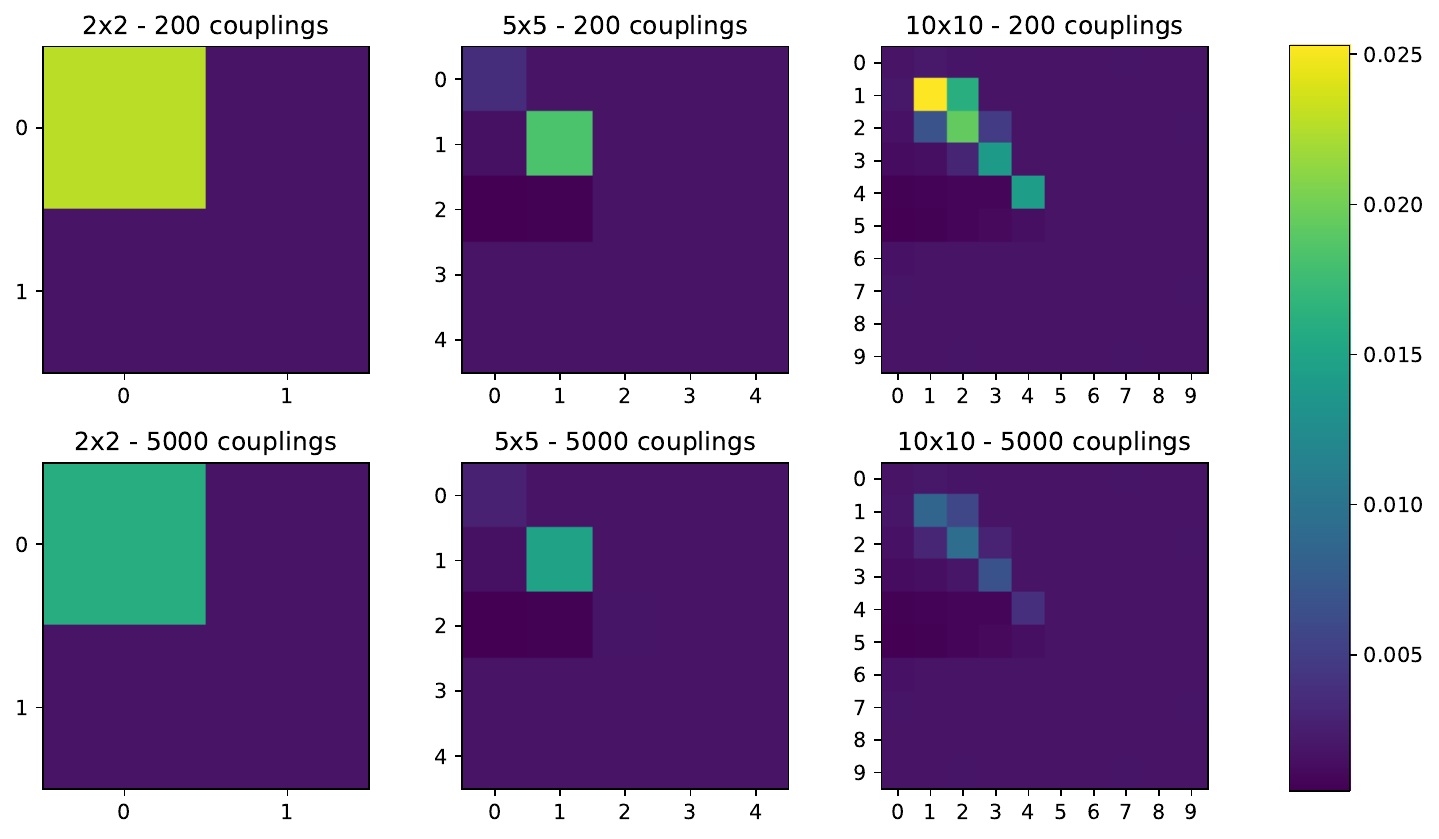}
    \caption{The simulated transmittances of 2×2, 5×5, and 10×10 test images, obtained using TURSCA with scattering couplings amounts of 200 and 5000, using the input parameters presented in Table \ref{tab:convg_simu_specs}.} 
    \label{fig:Trans_Heatmap}
\end{figure}

\begin{figure}[h]
    \centering   
    \includegraphics[width=1\textwidth]{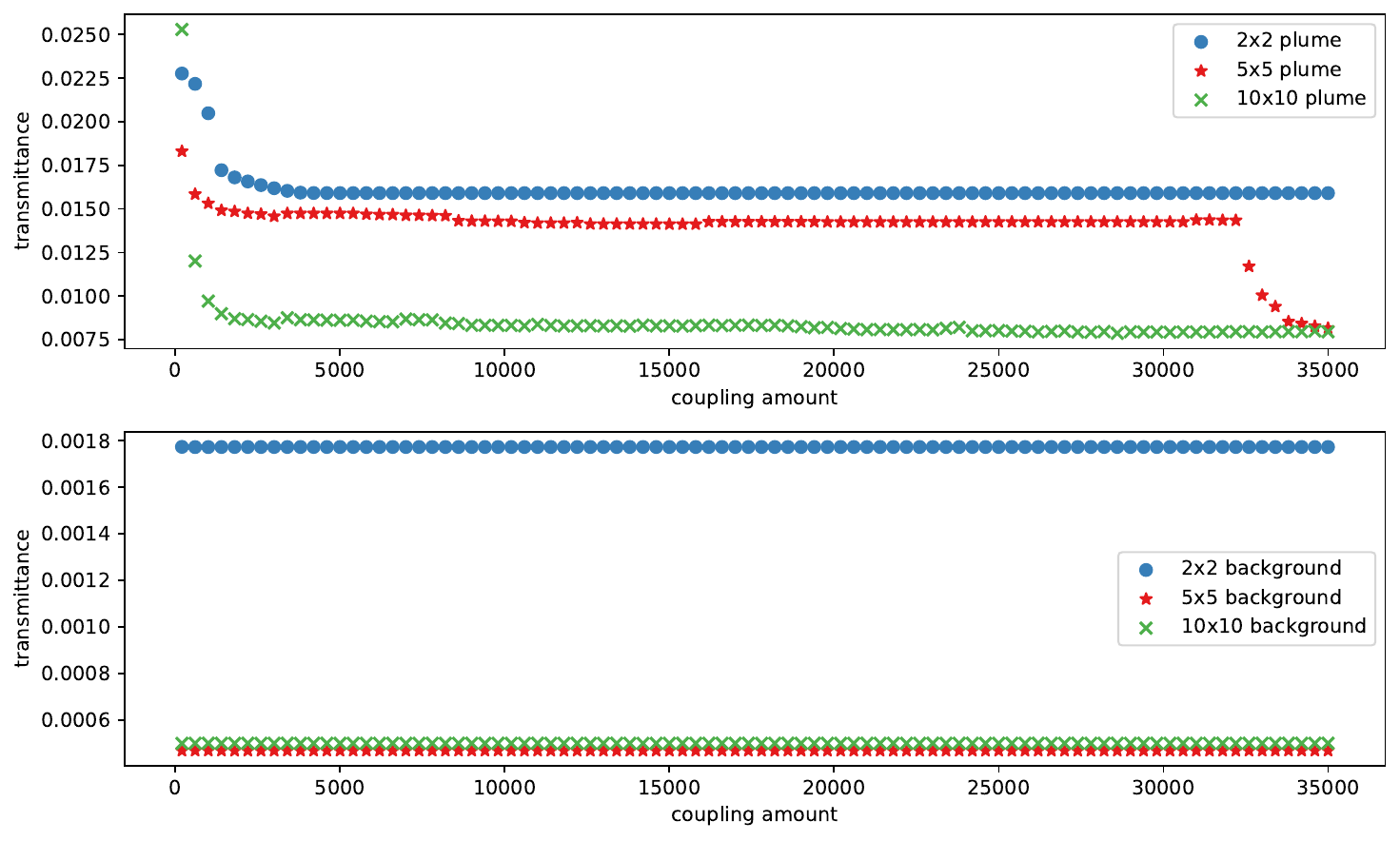}
    \caption{Transmittances as a function of coupling amount in the plume and background pixels. The maximum-value pixels were chosen to represent the plume, while the minimum-value pixels represent the background. The corresponding pixels are shown in Fig. \ref{fig:Trans_Heatmap}. The coordinates(row,col) of the maximum pixels for the 2x2, 5x5, and 10x10 are (0,0), (1,1), and (1,1), respectively, while the coordinates of the minimum pixels are (1,0), (2,0), and (5,0), respectively.} 
    \label{fig:Couplings-Transmittances}
\end{figure}

The TURSCA transmittances was examined using a nadir-viewing simulated scene of an emission plume at 1.99 $\mu$m wavelength. The emission source is situated in the middle of the scene and it emitted 410 kg/s of carbon, 98\% of which was black carbon aerosols by mass and the rest were CO$_2$. The carbon emission is carried by a 5 m/s wind toward north-west. Background atmosphere is Rayleigh scattering and contains absorbing CO$_2$ and H$_2$O, whose profiles are available on the previously referred TURSCA Github repository. The plume is composed of 30 overlaid spherical Gaussian basis functions with standard deviation ranging from 0.1 km at the emission source to 0.4 km at the plume tail end. A 30$\times$30 image single-scattering transmittance is presented in Fig. \ref{fig:plume_ss_30}. The specifics of the simulation parameters are presented in Table \ref{tab:convg_simu_specs}.
\begin{table}[]
    \centering
    \caption{Specifics of TURSCA simulations for convergence analysis.}
    \begin{tabular}{cc}
    \hline Input parameter & Value \\
       \hline Solar zenith angle & 77 degrees \\
       Solar azimuth angle & 56 degrees clockwise from image north\\
       Wavelength & 1.99 µm\\
       Lambertian surface albedo & 0.1 \\
       Surface pressure & 1.0139 atm \\
       Instrument altitude & 417 km \\
       Absorbing gases & CO$_2$, H$_2$O\\
       Aerosol radius & 0.05 $\mu$m\\
       Aerosol refractive index & $1.92 - 0.348i$ \\
       Avg. cosine of phase function & 0.006\\
       Field-of-view & $0.82^\circ\;\times\;0.82^\circ$\\
       \hline
    \end{tabular}
    \label{tab:convg_simu_specs}
\end{table}
The transmittances as a function of the amount of scattering coupling was analyzed with 2$\times$2, 5$\times$5 and 10$\times$10 images with the same field-of-view as Fig. \ref{fig:plume_ss_30}. Example simulation results are presented in Fig. \ref{fig:Trans_Heatmap}. The coupling amounts used in this analysis ranged from 0 to 35000, with the minimum and maximum pixels chosen to represent the background and plume, respectively. The results of this analysis are shown in Fig. \ref{fig:Couplings-Transmittances}.

As seen in Fig. \ref{fig:Couplings-Transmittances}, the transmittances of the chosen background pixel are nearly constant with respect to the coupling amount. This result is consistent with the other low-valued pixels, which were omitted from this figure for clarity reasons. Initially, in the plume pixel, up to 30000 couplings, it appeared that the transmittances would stabilize after couple thousand couplings. However, as observed in the 5×5 case, around 33000 couplings, there is a noticeable drop in simulated transmittances. Similar behaviour could occur in the other cases as well as the amount of scattering couplings increases. By examining the couplings, it was uncovered that this drop phenomenon was caused by a sudden increase of couplings within, to and from the plume pixel, which caused the couplings to be created unevenly across the image, resulting a potentially unphysical result.

These tests indicate that increasing the coupling amount will not always yield better results. More research is needed for finding the optimal coupling amount and the scattering coupling function (Eq. \ref{eq:scatt_coup}).

\begin{figure}
    \centering
    \includegraphics[width=0.9\linewidth]{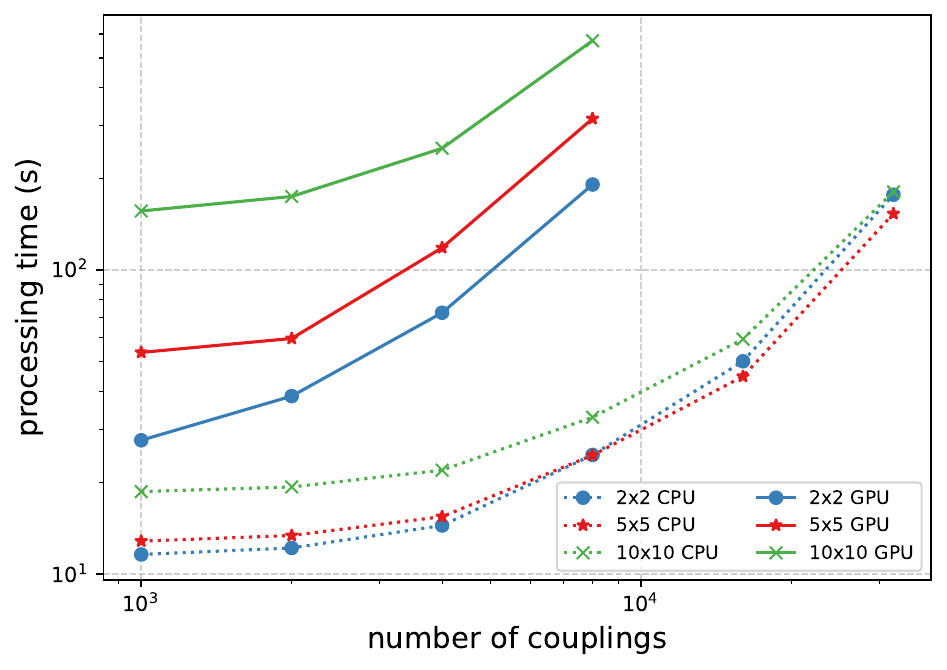}
    \caption{The runtimes of the TURSCA model on Intel Core i5-12600K CPU and on Asus GeForce RTX 3090 GPU by scattering coupling amount for image sizes 2$\times$2, 5$\times$5 and 10$\times$10 of the emission plume scene.}
    \label{fig:tursca-runtimes}
\end{figure}

Additionally, processing times were also studied, and the results can be found in Fig. \ref{fig:tursca-runtimes}. However, since the scattering graph method is a novel approach, runtime optimization of the computer code implementation has not been a top priority in this work and therefore calculation time superiority compared to other methods should not be expected. The runtime tests were conducted on Intel Core i5-12600K CPU and on Asus GeForce RTX 3090 GPU for image sizes 2$\times$2, 5$\times$5 and 10$\times$10, following the similar approach to the transmittance tests. The result suggests that runtime is more affected by the coupling amount rather than the image size. This effect is especially apparent in the CPU results, though it seems that the 2$\times$2 image experiences a slightly faster increase with the amount of couplings. This could be due to the smallest image approaching to the theoretical maximum number of scattering couplings and therefore the inter-node Beer-Lambert-Bouguer transmittance calculations take longer than in other cases.

Surprisingly the CPU is much faster than the GPU in all situations. However, the TURSCA algorithm in its current form has a couple of serialized sections which can cause a highly constraining performance bottleneck on the GPU. Also, the memory access patterns were designed with CPU in mind, which might be unoptimal for GPU computation. These results underline the importance of further algorithmic optimization work.

\section{Conclusions}
The scattering graph method was presented and its computational implementation TURSCA was described. TURSCA was compared against two established RT models and it was shown to be adequately accurate in the examined cases, with about 3\% difference against Siro and 6\% difference against SHDOM. The transmittances as a function of the amount of scattering coupling were examined: the background pixels showed little variation, while the plume case revealed a complex relationship with scattering couplings. Determining the optimal amount of scattering coupling and the scattering coupling function requires further research.  

This new approach to 3D RT modeling could enable faster and more accurate atmospheric remote sensing retrieval algorithms, since contrary to 1D RT models more traditionally found in atmospheric retrieval algorithms, scattering graph method simulates the RT of several pixels at once, which enables the utilization of physics-based inter-pixel correlations. Scattering graph method would also enable the analysis of sub-pixel inhomogeneities, which cannot be fully modeled using 1D RT. The method could be especially useful for greenhouse gas retrieval algorithms, which are sensitive to measurement noise and require lots of auxiliary information.

The current study enables thorough analysis of the scattering graph method. Some of the further work include a review of the effects of different coupling functions, the examination of line-of-sight scattering node dotting procedure, and a strategy for finding the optimal coupling threshold. Additionally, scattering graph partitioning should be looked into so that the computation could be further parallelized.

The TURSCA implementation also requires further refinement. Extensive benchmarking of each part of the code is needed for finding performance bottlenecks on both CPUs and GPUs. The model capabilities could be expanded to polarized, spectral and thermal RT. Depending on the desired use case, the medium basis functions, boundary shape and reflectance functions as well as light source definitions can be readily expanded.

\begin{backmatter}
\bmsection{Funding}
The research has been partially funded by the Research Council of Finland (grant agreements 337552, 331829, 351311, 359196, 353982) and Centre for Scientific Computation FICOCOSS project (grant agreement CNTR0015154).
%Content in the funding section will be generated entirely from details submitted to Prism. Authors may add placeholder text in this section to assess length, but any text added to this section will be replaced during production and will display official funder names along with any grant numbers provided. If additional details about a funder are required, they may be added to the Acknowledgment, even if this duplicates some information in the funding section. For preprint submissions, please include funder names and grant numbers in the manuscript.

\bmsection{Acknowledgment}
A. Mikkonen thanks Jesse Loveridge of Colorado State University for consultation with running the SHDOM software. A. Mikkonen thanks Vijay Natraj of NASA Jet Propulsion Laboratory for fruitful discussions regarding atmospheric radiative transfer.

\bmsection{Disclosures}
The authors declare no conflicts of interest.

\bmsection{Data Availability Statement}
Data underlying the results presented in this paper are available in Ref. \cite{tursca-github}.

\end{backmatter}

%\section{References}
%\label{sec:refs}

%%%%%%%%%%%%%%%%%%%%%%% References %%%%%%%%%%%%%%%%%%%%%%%%%

%%%%%%%%%% If using BibTeX:
\bibliography{bibliography}

\begin{thebibliography}{10}
\newcommand{\enquote}[1]{``#1''}

\bibitem{amt-5-99-2012-odell}
C.~W. O'Dell, B.~Connor, H.~B\"osch, \emph{et~al.}, \enquote{The {ACOS CO}$_{2}$ retrieval algorithm – {Part} 1: Description and validation against synthetic observations,} {\protect\JournalTitle{Atmospheric Measurement Techniques}} \textbf{5}, 99--121 (2012).

\bibitem{spurr2008lidort}
R.~Spurr, \enquote{{LIDORT and VLIDORT}: Linearized pseudo-spherical scalar and vector discrete ordinate radiative transfer models for use in remote sensing retrieval problems,} {\protect\JournalTitle{Light scattering reviews 3: Light scattering and reflection}} pp. 229--275 (2008).

\bibitem{amt-9-5423-2016-hu}
H.~Hu, O.~Hasekamp, A.~Butz, \emph{et~al.}, \enquote{The operational methane retrieval algorithm for {TROPOMI},} {\protect\JournalTitle{Atmospheric Measurement Techniques}} \textbf{9}, 5423--5440 (2016).

\bibitem{SCHEPERS2014347}
D.~Schepers, J.~{aan de Brugh}, P.~Hahne, \emph{et~al.}, \enquote{{LINTRAN} v2.0: A linearised vector radiative transfer model for efficient simulation of satellite-born nadir-viewing reflection measurements of cloudy atmospheres,} {\protect\JournalTitle{Journal of Quantitative Spectroscopy and Radiative Transfer}} \textbf{149}, 347--359 (2014).

\bibitem{amt-18-73-2025-trees}
V.~J.~H. Trees, P.~Wang, P.~Stammes, \emph{et~al.}, \enquote{Cancellation of cloud shadow effects in the absorbing aerosol index retrieval algorithm of {TROPOMI},} {\protect\JournalTitle{Atmospheric Measurement Techniques}} \textbf{18}, 73--91 (2025).

\bibitem{amt-16-2145-2023}
S.~T. Massie, H.~Cronk, A.~Merrelli, \emph{et~al.}, \enquote{Insights into {3D} cloud radiative transfer effects for the {Orbiting Carbon Observatory},} {\protect\JournalTitle{Atmospheric Measurement Techniques}} \textbf{16}, 2145--2166 (2023).

\bibitem{amt-14-2127-2021-jervis}
D.~Jervis, J.~McKeever, B.~O.~A. Durak, \emph{et~al.}, \enquote{The {GHGSat-D} imaging spectrometer,} {\protect\JournalTitle{Atmospheric Measurement Techniques}} \textbf{14}, 2127--2140 (2021).

\bibitem{nasa-emit}
R.~O. Green, N.~Mahowald, C.~Ung, \emph{et~al.}, \enquote{The earth surface mineral dust source investigation: An earth science imaging spectroscopy mission,} in \emph{2020 IEEE Aerospace Conference,}  (2020), pp. 1--15.

\bibitem{8518512-prisma}
R.~Loizzo, R.~Guarini, F.~Longo, \emph{et~al.}, \enquote{{Prisma: The Italian Hyperspectral Mission},} in \emph{IGARSS 2018 - 2018 IEEE International Geoscience and Remote Sensing Symposium,}  (2018), pp. 175--178.

\bibitem{10.1117/12.2304099-sentinel4}
S.~T. Gulde, M.~G. Kolm, D.~J. Smith, \emph{et~al.}, \enquote{{Sentinel 4: a geostationary imaging UVN spectrometer for air quality monitoring: status of design, performance and development},} in \emph{International Conference on Space Optics — ICSO 2014,}  vol. 10563 Z.~Sodnik, B.~Cugny, and N.~Karafolas, eds., International Society for Optics and Photonics (SPIE, 2017), p. 1056341.

\bibitem{pincus-2009-mc-shdom}
R.~Pincus and K.~F. Evans, \enquote{{Computational Cost and Accuracy in Calculating Three-Dimensional Radiative Transfer: Results for New Implementations of Monte Carlo and SHDOM},} {\protect\JournalTitle{Journal of the Atmospheric Sciences}} \textbf{66}, 3131 -- 3146 (2009).

\bibitem{jones-2018-mc-shdom}
A.~L. Jones and L.~D. Girolamo, \enquote{{Design and Verification of a New Monochromatic Thermal Emission Component for the I3RC Community Monte Carlo Model},} {\protect\JournalTitle{Journal of the Atmospheric Sciences}} \textbf{75}, 885 -- 906 (2018).

\bibitem{i3rc}
R.~F. Cahalan, L.~Oreopoulos, A.~Marshak, \emph{et~al.}, \enquote{{THE I3RC: Bringing Together the Most Advanced Radiative Transfer Tools for Cloudy Atmospheres},} {\protect\JournalTitle{Bulletin of the American Meteorological Society}} \textbf{86}, 1275 -- 1294 (2005).

\bibitem{shdom}
K.~F. Evans, \enquote{The spherical harmonics discrete ordinate method for three-dimensional atmospheric radiative transfer,} {\protect\JournalTitle{Journal of the Atmospheric Sciences}} \textbf{55}, 429 -- 446 (1998).

\bibitem{GRAU2013149-DART}
E.~Grau and J.-P. Gastellu-Etchegorry, \enquote{{Radiative transfer modeling in the Earth–Atmosphere system with DART model},} {\protect\JournalTitle{Remote Sensing of Environment}} \textbf{139}, 149--170 (2013).

\bibitem{govaerts1998raytran}
Y.~M. Govaerts and M.~M. Verstraete, \enquote{{Raytran: A Monte Carlo ray-tracing model to compute light scattering in three-dimensional heterogeneous media},} {\protect\JournalTitle{IEEE Transactions on geoscience and remote sensing}} \textbf{36}, 493--505 (1998).

\bibitem{QI2019695-LESS}
J.~Qi, D.~Xie, T.~Yin, \emph{et~al.}, \enquote{{LESS: LargE-Scale remote sensing data and image simulation framework over heterogeneous 3D scenes},} {\protect\JournalTitle{Remote Sensing of Environment}} \textbf{221}, 695--706 (2019).

\bibitem{luebke-cuda-2008}
D.~Luebke, \enquote{{CUDA: Scalable parallel programming for high-performance scientific computing},} in \emph{2008 5th IEEE International Symposium on Biomedical Imaging: From Nano to Macro,}  (2008), pp. 836--838.

\bibitem{munshi-opencl-2009}
A.~Munshi, \enquote{{The OpenCL specification},} in \emph{2009 IEEE Hot Chips 21 Symposium (HCS),}  (2009), pp. 1--314.

\bibitem{RAMON201989}
D.~Ramon, F.~Steinmetz, D.~Jolivet, \emph{et~al.}, \enquote{{Modeling polarized radiative transfer in the ocean-atmosphere system with the GPU-accelerated SMART-G Monte Carlo code},} {\protect\JournalTitle{Journal of Quantitative Spectroscopy and Radiative Transfer}} \textbf{222-223}, 89--107 (2019).

\bibitem{HUANG20112207}
B.~Huang, J.~Mielikainen, H.~Oh, and H.-L. {Allen Huang}, \enquote{{Development of a GPU-based high-performance radiative transfer model for the Infrared Atmospheric Sounding Interferometer (IASI)},} {\protect\JournalTitle{Journal of Computational Physics}} \textbf{230}, 2207--2221 (2011).

\bibitem{bian-2022-3drtmc}
Z.~Bian, J.~Qi, J.~P. Gastellu-Etchegorry, \emph{et~al.}, \enquote{{A GPU-Based Solution for Ray Tracing 3-D Radiative Transfer Model for Optical and Thermal Images},} {\protect\JournalTitle{IEEE Geoscience and Remote Sensing Letters}} \textbf{19}, 1--5 (2022).

\bibitem{alerstam-2008}
E.~Alerstam, T.~Svensson, and S.~Andersson-Engels, \enquote{{Parallel computing with graphics processing units for high-speed Monte Carlo simulation of photon migration},} {\protect\JournalTitle{Journal of Biomedical Optics}} \textbf{13}, 060504 (2008).

\bibitem{EFREMENKO20143079}
D.~S. Efremenko, D.~G. Loyola, A.~Doicu, and R.~J. Spurr, \enquote{{Multi-core-CPU and GPU-accelerated radiative transfer models based on the discrete ordinate method},} {\protect\JournalTitle{Computer Physics Communications}} \textbf{185}, 3079--3089 (2014).

\bibitem{ishimaru1978wave}
A.~Ishimaru, \emph{Wave propagation and scattering in random media}, vol.~1 (Academic Press, 1978).

\bibitem{MIKKONEN2024108892}
A.~Mikkonen, H.~Lindqvist, J.~Peltoniemi, and J.~Tamminen, \enquote{{Non-Lambertian snow surface reflection models for simulated top-of-the-atmosphere radiances in the NIR and SWIR wavelengths},} {\protect\JournalTitle{Journal of Quantitative Spectroscopy and Radiative Transfer}} \textbf{315}, 108892 (2024).

\bibitem{gonzalez2010measurement}
{\'A}.~Gonz{\'a}lez, \enquote{{Measurement of areas on a sphere using Fibonacci} and latitude--longitude lattices,} {\protect\JournalTitle{Mathematical geosciences}} \textbf{42}, 49--64 (2010).

\bibitem{manalo2015advanced}
K.~Manalo, C.~D. Ahrens, and G.~Sjoden, \enquote{Advanced quadratures for three-dimensional discrete ordinate transport simulations: A comparative study,} {\protect\JournalTitle{Annals of Nuclear Energy}} \textbf{81}, 196--206 (2015).

\bibitem{taichi-2019}
Y.~Hu, T.-M. Li, L.~Anderson, \emph{et~al.}, \enquote{Taichi: a language for high-performance computation on spatially sparse data structures,} {\protect\JournalTitle{ACM Trans. Graph.}} \textbf{38} (2019).

\bibitem{oikarinen1999multiple}
L.~Oikarinen, E.~Sihvola, and E.~Kyr{\"o}l{\"a}, \enquote{Multiple scattering radiance in limb-viewing geometry,} {\protect\JournalTitle{Journal of Geophysical Research: Atmospheres}} \textbf{104}, 31261--31274 (1999).

\bibitem{zawada2021systematic}
D.~Zawada, G.~Franssens, R.~Loughman, \emph{et~al.}, \enquote{Systematic comparison of vectorial spherical radiative transfer models in limb scattering geometry,} {\protect\JournalTitle{Atmospheric Measurement Techniques}} \textbf{14}, 3953--3972 (2021).

\bibitem{rs12172831}
A.~Levis, Y.~Y. Schechner, A.~B. Davis, and J.~Loveridge, \enquote{Multi-view polarimetric scattering cloud tomography and retrieval of droplet size,} {\protect\JournalTitle{Remote Sensing}} \textbf{12} (2020).

\bibitem{ZHANG2022104054}
D.~Zhang, L.~Bai, Y.~Wang, \emph{et~al.}, \enquote{{An improved SHDOM coupled with CFD for simulating infrared radiation signatures of rocket plumes},} {\protect\JournalTitle{Infrared Physics \& Technology}} \textbf{122}, 104054 (2022).

\bibitem{DOICU2021107386}
A.~Doicu, M.~I. Mishchenko, D.~S. Efremenko, and T.~Trautmann, \enquote{Spectral spherical harmonics discrete ordinate method,} {\protect\JournalTitle{Journal of Quantitative Spectroscopy and Radiative Transfer}} \textbf{258}, 107386 (2021).

\bibitem{tursca-github}
\enquote{{GitHub} - amikko/tursca: {Monochromatic scalar 3D radiative transfer model TURSCA},} \url{https://github.com/amikko/tursca}. Accessed: 2025-07-23.

\end{thebibliography}

%%%%%%%%%% If preparing manually:
% \begin{thebibliography}{1}
% \newcommand{\enquote}[1]{``#1''}

% \bibitem{Zhang:14}
% Y.~Zhang, S.~Qiao, L.~Sun, Q.~W. Shi, W.~Huang, L.~Li, and Z.~Yang,
%   \enquote{Photoinduced active terahertz metamaterials with nanostructured
%   vanadium dioxide film deposited by sol-gel method,}
%   {\protect\JournalTitle{Optics Express}} \textbf{22}, 11070--11078 (2014).

% \bibitem{Optica}
% {Optica}, \enquote{{Optica Publishing Group},}
%   \url{http://www.opg.optica.org}.

% \bibitem{FORSTER2007}
% P.~Forster, V.~Ramaswamy, P.~Artaxo, T.~Bernsten, R.~Betts, D.~Fahey,
%   J.~Haywood, J.~Lean, D.~Lowe, G.~Myhre, J.~Nganga, R.~Prinn, G.~Raga,
%   M.~Schulz, and R.~V. Dorland, \enquote{Changes in atmospheric consituents and
%   in radiative forcing,} in \enquote{Climate Change 2007: The Physical Science
%   Basis. Contribution of Working Group 1 to the Fourth Assesment Report of
%   Intergovernmental Panel on Climate Change,}  S.~Solomon, D.~Qin, M.~Manning,
%   Z.~Chen, M.~Marquis, K.~B. Averyt, M.~Tignor, and H.~L. Miler, eds.
%   (Cambridge University Press, 2007).

% \end{thebibliography}

\end{document}